\documentclass[aps,twocolumn,showpacs,superscriptaddress]{revtex4-1}
\usepackage{amsfonts}
\usepackage{amsmath}
\usepackage{amssymb}
\usepackage{lipsum}
\usepackage{graphicx}
\usepackage{bm}
\usepackage{natbib}
\usepackage{float}
\usepackage[dvipsnames]{xcolor}
\newcommand{\sgn}{\rm sign}

\begin{document}
\preprint{APS/123-QED}
\title{
Local magnetic moments and 
electronic transport 
in 
closed loop quantum dot systems: a case of quadruple quantum dot ring at and away from equilibrium
}
\author{V. S. Protsenko}
\affiliation{%
M. N. Mikheev Institute of Metal Physics, 620990 Ekaterinburg Russia}
\affiliation{
Ural Federal University, 620002 Ekaterinburg, Russia
}
\author{A. A. Katanin}%
\affiliation{%
M. N. Mikheev Institute of Metal Physics, 620990 Ekaterinburg Russia}
\date{\today}
\begin{abstract}
We apply the non-equilibrium functional renormalization group approach treating flow of the electronic self-energies, to describe
local magnetic moments formation and electronic transport in a quadruple quantum dot (QQD) ring, coupled to leads, with moderate Coulomb interaction on the quantum dots. 
We find that at zero temperature depending on parameters of the QQD system the regimes with zero, one, or two almost local magnetic moments in the ring can be realized, and the results of the considered approach in equilibrium agree qualitatively with those of more sophisticated fRG approach treating also flow of the vertices. It is shown that the almost formed local magnetic moments, which exist in the equilibrium,
remain stable in a wide range of bias voltages near equilibrium.
The destruction of the local magnetic moments with increasing bias voltage 
is realized in one or two stages, depending on the parameters of the system; for two-stage process the intermediate phase possesses fractional magnetic moment. We present zero-temperature results for current-voltage dependences and differential conductances of the system, which exhibit sharp features at the transition points between different magnetic states. The occurrence of interaction induced negative differential conductance phenomenon is demonstrated and discussed. For one local moment in the ring and finite hopping between the opposite quantum dots, connected to the leads, we find suppression of the conductance for one of the spin projections in infinitesimally small magnetic field, which occurs due to destructive interference of different electron propagation paths and can be used in spintronic devices. 
\end{abstract}
\maketitle
\section{Introduction}\par
Quantum dots are nano-scale crystals, which have a discrete energy spectrum and for this reason often referred to as artificial atoms. Systems based on quantum dots are potentially important for nanoelectronic 
applications. 
Due to different kinds of topologies of these systems, the multi-dot systems 
can show non-trivial interplay of fundamental effects (e.g., Fano and Kondo effects~\cite{Tanamoto_2007,Jiang_2008}).
An especially rich physics emerges when the geometry of the system allows electron tunneling through closed-loop geometries. Many fundamental effects such as Fano resonances~\cite{Hackenbroich_2001,Guevara_2006,Meden_2006,Zeng_2002}, Aharonov-Bohm oscillations~\cite{Loss_2000,Chi_2006,Delgado_2008,Zeng_2002}, Kondo behavior~\cite{Shang_2015,Oguri_2009,Tanamoto_2007,Aguado_2000} and corresponding quantum phase transitions (QPT) ~\cite{Zitko_2007(2),Wang_2007,Zitko_2008,Tooski_2016,Liu_2010,Zitko_2006,Dias_2006,ZH}  have been found in these systems. 

The simplest system of this kind is the parallel double quantum dot (DQD) system \cite{Zitko_2007(2),Zitko_2006,Dias_2006,Wang_2007,ZH,PK_2016,PK_2017,Chi_2005,Meden_2006,Trocha_2007,Tanaka_2005,Holleitner_2001}. It was shown that this 
system even for moderate values of Coulomb interaction may demonstrate the interaction-induced QPTs to the so-called singular Fermi liquid (SFL) state, associated with the presence of the local magnetic moment in one of the states (so called "odd" state), which is weakly hybridized or decoupled from the conduction bands (leads).
The SFL state remains stable in a wide range of gate voltages near half filling and at some critical gate voltage undergoes QPT into paramagnetic state without local moments.
The type of the QPTs and peculiarities of the electron transport at the transition strongly depend on the type of the system symmetry, as well as on the number of energy levels. In particular, for the parallel double quantum dot system it was found that depending on the 
symmetry of the system it can demonstrate either a first-order QPT to SFL state, accompanied by a discontinuous change of the conductance or the second-order QPT, in which the conductance is continuous and exhibits Fano-type asymmetric resonance near the transition point~\cite{PK_2017}. In both cases, the conductance reaches almost unitary limit in the SFL phase. Therefore, the QPTs to SFL state have a significant impact on the electron transport.

The SFL state may occur 
also in other closed loop geometries of atoms or quantum dots, appearing in larger nanoscopic systems, e.g.~organic molecules~\cite{Ke_2008,Markussen_2010,Cardamone_2006,Guedon_2012,Quian_2008}, quantum networks~\cite{Foldi_2008,Dey_2011,Fu_2012} etc., where the interference of different paths of electron propagation 
may yield non-trivial quantum phase transitions and transport properties.
 The electron-electron interaction plays an important role in these systems. At the same time, numerically exact methods such as the numerical renormalization group, experience serious difficulties for large number of interacting sites. 

As a simplest multi-dot system with closed loop geometry, in the present paper we study the quadruple quantum dot (QQD) ring system
\cite{Zeng_2002,Shang_2015,Liu_2010,
Lovey_2011,Yan_2006,Ex1,Ex2}. This system appears as a building block of quantum network devices \cite{Foldi_2008,Dey_2011,Fu_2012}. This system can be also viewed as a prototype of cyclobutadiene organic molecule, discussed some time ago from the viewpoint of electronic transport \cite{cybtdn}.
The QQD system 
demonstrates a rather rich phase diagram with the possibility of controlling 
spin states of electrons 
\cite{Ozfidan_2013}, making 
it promising for the development of spintronic devices. 
It was shown that spin-polarized electron transport ~\cite{Fu_2012,Kagan_2017,Wu_2013,Eslami_2014}, e.g. generated by tuning the energy levels or hopping
levels in an external magnetic field~\cite{Fu_2012,Kagan_2017}, can be achieved in this structure. 

Various spin states found for the isolated QQD system \cite{Ozfidan_2013} 
imply a possibility of different magnetic moment states in this system connected to the leads
even in the absence  of the (or in the infinitesimal) magnetic field. 
In this respect, study of the possibility of the formation 
of spin-split (in vanishingly small magnetic field) phases, corresponding to presence of local magnetic moments, 
their connection with the transport properties of the system, and evolution under the non-equilibrium conditions 
opens a way to model larger systems, including quantum networks and organic molecules. 
Although paramagnetic solution self-energies of QQD were studied in \cite{DGAParquet,Kagan_2017,KA2}, they do not give sufficient information on the formation of local magnetic moments. 

Only a limited number of studies have been done on the non-equilibrium effects in the QQD systems and mainly focused on the effects of the spin-polarization, magnification and circulation of the persistent current~\cite{Yi_2012,Yi_2010}, as well the current oscillation phenomena. The current-voltage $(J-V)$ characteristics have been investigated in some particular cases~\cite{Wu_2014,Wu_2013,KA2}, including a possibility of 
negative differential conductance (NDC) effects~\cite{KA2}, analogous to those previously found for parallel quantum dots \cite{Aguado_2000,Lara_2008,Fransson_2004,Liu_2007,Chi_2005}. These studies however did not investigate in detail the possibility and effects of local moment formation, e.g. away from equilibrium. 
From practical point of view, it is also interesting to consider whether it is possible to obtain highly spin-polarized current due to the energy difference of the spin-up and -down states, caused by the transition to the magnetic moment state in an infinitesimal magnetic field without the spin-orbit interaction.\par


To study the above mentioned aspects of electronic and transport properties of QQD system we use the functional renormalization-group approach 
~\cite{Karrasch_2006,Metzner_2012,Gezzi_2007,Jakobs_2007}. This approach (after introducing the appropriate counterterm, which corresponds to switching off or decreasing magnetic field during the flow) was able to describe both, normal and SFL phases of the DQD system and was found to be in a good agreement with the numerical renormalization group data for a parallel quantum dot system in equilibrium up to intermediate values of the Coulomb interaction~\cite{PK_2016,PK_2017}. 
However, generalization of this approach to larger systems is not straightforward, since it yields  electron interaction vertices, which number increases as 4-th power of number of quantum dots, which are also difficult to treat numerically. 

In the present paper we exploit the fRG method, which neglects flow of the electron interaction vertices, to describe one of the simplest systems of quantum dots, forming closed loop. The considered method represents a generalization of the fRG approach~\cite{Karrasch_2006,Metzner_2012} to the Keldysh space \cite{Gezzi_2007,Jakobs_2007} and allows one to reformulate an interacting problem in terms of coupled differential equations for flowing self-energies, which, after several approximations, can be easily integrated even for complex systems. 
 Among 
other methods, the non-equilibrium fRG approach has some advantages: it does not require significant computational resources and results of its implementation 
are consistent with the ones obtained through more elaborate methods deal with non-equilibrium situations~\cite{Karrasch_2010}. 
Recently, this method has been successfully applied to several quantum dot systems~\cite{Karrasch_2010,Karrasch_2010_2,Jakobs_2010,Kennes_2013,Rentrop_2014} and the comparative study to other numerical and semi-analytical methods has been done~\cite{Karrasch_2010,Eckel_2010}. Its application to systems with closed loop geometries formed by quantum dots was not however performed so far. 

We argue that the considered method is able to describe various aspects of electronic properties
of quantum dot or molecular systems with closed loop geometries, which are exemplified by QQD system. 
In particular, we consider both, equilibrium and non-equilibrium regimes of the QQD system in the zero-temperature limit $T=0$.
In equilibrium, we show that depending on the geometry of the QQD system the regimes with zero, one, and two almost local magnetic moments can be realized. Moreover, adding hopping between the opposite quantum dots, attached to the contacts, allows one to use this system as a spin filter even in the absence of the spin-orbit coupling: for sufficiently large hopping in a certain range of gate voltages we find zero conductance for one of the spin projections (oriented {along} the infinitesimally small magnetic field). 

We find that the  magnetic moments, existing at zero bias voltage, remain stable in the wide range of bias voltages near equilibrium. At the same time, at higher bias voltages the destruction of the magnetic moments occurs and proceeds in one or two stages, depending on the parameters of the QQD system. We present results for the current-voltage characteristic and the differential conductance of the system, which exhibit sharp features at the transition points between different magnetic phases. The occurrence of interaction induced NDC phenomena is demonstrated. The presented method may be therefore used to describe electronic transport in larger systems: quantum networks and organic molecules. \par

The paper is organized as follows. In Sect. II we introduce the model and briefly discuss the non-equilibrium fRG method.  In Sect. III we present the results of the fRG calculations in equilibrium and analyze the possibility of the local moments formation (Sect. IIIA) and differential conductance (Sects. IIIB,C). In Sect. IVA we discuss non-equilibrium regime, and in Sect. IVB we present the $J$-$V$ characteristics of the QQD system and discuss the appearance of the NDC phenomenon. Finally, in Sect. V we present conclusions. 
\section{Model and method}
\begin{figure}[t]
\center{\includegraphics[width=0.75\linewidth]{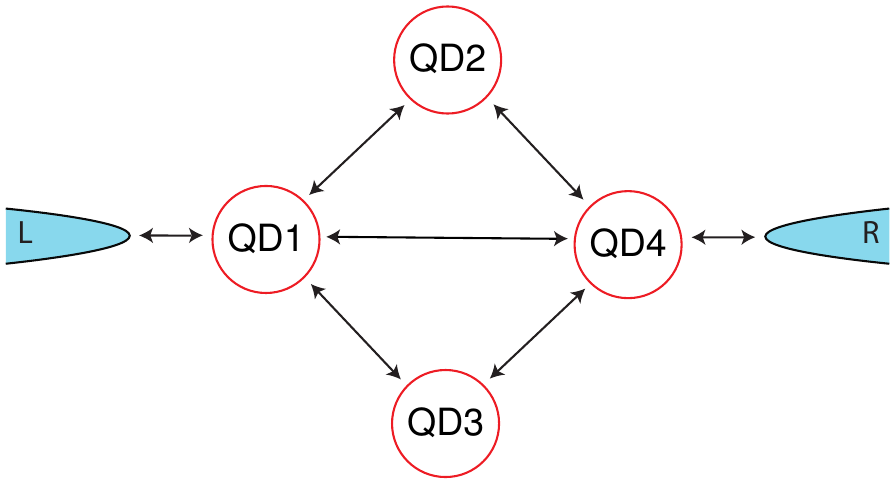}}
\caption{(Color online). Schematic representation of quadruple quantum dot structure (QD1-QD4) connected to two left (L) and right (R) leads.}
\label{sketch}
\end{figure}

We consider the QQD system as depicted in Fig. \ref{sketch}. The corresponding model is defined by the following Hamiltonian
\begin{equation}
 \mathcal{H}= \mathcal{H}_{\rm QQD}+\mathcal{H}_{\rm leads}+\mathcal{H}_{\rm T}.
 \label{Hamiltonian}
\end{equation}
The term $\mathcal{H}_{\rm QQD}$ in the Eq.~(\ref{Hamiltonian}) describes the isolated QQD cluster, 
\begin{eqnarray}
 \mathcal{H}_{\rm QQD}&=&\sum_{\sigma}\sum_{j=1}^{4}\left[\left(\epsilon_{j}-\sigma H-
 U_{j}/2\right)d^{\dagger}_{j,\sigma}d_{j,\sigma}\right.\notag\\&+&
 (U_{j}/2)n_{j,\sigma}n_{j,\bar{\sigma}}\Big]-\sum_{\sigma}{\left[\left(t_{12}d^{\dagger}_{1,\sigma}d_{2,\sigma}\right.\right.}\notag\\&+&{\left.\left.
t_{24}d^{\dagger}_{2,\sigma}d_{4,\sigma}+t_{13}d^{\dagger}_{1,\sigma}d_{3,\sigma}+t_{34}d^{\dagger}_{3,\sigma}d_{4,\sigma}\right.\right.}\notag\\
&+&{\left.\left.t_{14}d^{\dagger}_{1,\sigma}d_{4,\sigma}\right)\right.}+\text{H.c.}\Big],
 \label{H_dot}
\end{eqnarray}
where $d^{\dagger}_{j,\sigma} (d_{j,\sigma})$ are the creation (annihilation) operators for electrons with spin $\sigma\in\{\uparrow(1/2),\downarrow(-1/2)\}$ $(\bar{\sigma}=-\sigma)$ on the $j$-th quantum dot, $n_{j,\sigma}=d^{\dagger}_{j,\sigma}d_{j,\sigma}$. The parameters $\epsilon_{j}$ are the level positions, $H$ is the magnetic field, which produces Zeeman splitting of the energy levels (we assume in the following that QQD structure is not affected by magnetic flux), $U_{j}$ and $t_{ij}$ denote the on-site Coulomb interaction of the $j$-th dot and tunnel matrix elements between the nearest-neighbor quantum dots, respectively. In the following we assume that the quantum dots are equal, hence $U_{j}=U$ and $\epsilon_{j}=\epsilon$.\par
The second $\mathcal{H}_{\rm leads}$ and third $\mathcal{H}_{\rm T}$ terms in Eq.~(\ref{Hamiltonian}) describe the noninteracting leads and the tunneling of electrons between the leads and dots, respectively,  
\begin{eqnarray}
\mathcal{H}_{\rm leads}&=&-\sum_{\alpha=L,R}\sum_{k=0}^{\infty}\sum_{\sigma}\left[\mu_{\alpha}c^{\dagger}_{\alpha,k,\sigma}c_{\alpha,k,\sigma}\right.\notag\\
&+&\left.\tau(c^{\dagger}_{\alpha,k+1,\sigma}c_{\alpha,k,\sigma}+\text{H.c.})\right],
\label{H_leads}\\
\mathcal{H}_{\rm T}&=&-\sum_{\sigma}\left[\left(t_{L}c^{\dagger}_{L,0,\sigma}d_{1,\sigma}+t_{R}c^{\dagger}_{R,0,\sigma}d_{4,\sigma}\right)\right.\notag\\&+&\left.\text{H.c.}\right],
\label{H_T}
\end{eqnarray}
where $c^{\dagger}_{\alpha,k,\sigma}(c_{\alpha,k,\sigma})$ is the corresponding creation (annihilation) operator for an electron on the $k$ lattice site of the left $\alpha=L$ or right $\alpha=R$ lead, $\tau$ denotes  nearest-neighbor hopping between the sites of the leads, $\mu_{\alpha}$ is the chemical potential and $t_{\alpha}$ is the dot-lead coupling matrix element.\par

In the absence of the electron-electron interaction $U$ for 
hopping symmetry  $t_{12}/t_{13}=t_{24}/t_{34}$ one of the states (the so called odd state), obtained by an appropriate canonical transformation of the states on QD2,3 to the even-odd basis (see Appendix A, cf. Ref.~\cite{PK_2017}),
can be completely disconnected from the other quantum dots (and, consequently, from the leads). Even in the presence of the Coulomb interaction, this state remains weakly hybridized with the leads, which yields formation of the local moment in that state in the vicinity of half filling ($\epsilon_j$=0), see Sect. IIIA below. 
In this respect, the QQD system {at $t_{14}=0$} is similar to the double quantum dot system, where the presence of the odd, weakly hybridized with the leads, state provides the possibility for formation of a correlation induced local magnetic moment in the system~\cite{Zitko_2007(2),ZH,PK_2016,PK_2017}. {However, as it will be shown in Sect. IIIC below, apart from the tunneling through the even energy level, which takes place in DQD system, in QQD system  
the resonant tunneling from QD1 to QD4 is possible. This difference becomes especially prominent when switching on $t_{14}$ hopping}, which will be also discussed in Sect. IIIC.

The simplest asymmetry, which allows one to focus on the effect of the (local) magnetic moment formation under equilibrium and non-equilibrium conditions and its influence on the electron transport, 
is the so called diagonal hopping asymmetry~\cite{PK_2017} $t_{12}=t_{34}=t$, $t_{13}=t_{24}=\gamma t$, where the parameter $\gamma$ varies from zero to unity. This choice of the geometry allows us to study the evolution of the system from the case of $\gamma=1$, when all hopping matrix elements are equal and the local moment is formed in the odd state in the equilibrium, to the case of $\gamma=0$ for which the system splits into the two subsystems, each of which hybridized to only one of the leads, and the local moments are present in both, even and odd states for small hybridization to the leads, or absent otherwise. We do not consider hopping between the QD2,3 because it does not change qualitatively conductivities for small hoppings, and for large hoppings simply destroy local moments (if they were present without this hopping) due to mixing of even- and odd states.

By using the Dyson equation and the projection technique the bare Green function of the system in the Keldysh space can be written as
\begin{equation}
\mathcal{G}=\begin{pmatrix}
\mathcal{G}^{--}&\mathcal{G}^{-+}\\
\mathcal{G}^{+-}&\mathcal{G}^{++}
\end{pmatrix}
=\left[\mathcal{G}^{-1}_{\rm dots}-\Sigma_{\rm bath}\right]^{-1},
\label{Gf}
\end{equation}
where
\begin{eqnarray}
\left[\mathcal{G}^{-1}_{\rm dots}\right]^{kk^{'}}_{jj^{'};\sigma}&=&-k\delta_{kk^{'}}\\
&\times&\begin{pmatrix}
\omega-\epsilon_{1,\sigma} & t_{12} & t_{13} & t_{14}\\
t_{12} & \omega-\epsilon_{2,\sigma} & 0 & t_{24}\\
t_{13} & 0 & \omega-\epsilon_{3,\sigma} & t_{34}\\
t_{14} & t_{24} & t_{34} & \omega-\epsilon_{4,\sigma} 
\end{pmatrix}_{jj^{'}}\notag
\end{eqnarray}
with $\epsilon_{j,\sigma}=\epsilon_{j}-\sigma H$ is the Green function of the isolated QQD cluster and
\begin{eqnarray}
&&\left[\Sigma_{\rm bath}\right]^{kk^{'}}_{jj^{'};\sigma}=-i\delta_{jj^{'}}\sum_{\alpha}\Theta_j^\alpha\Gamma_{\alpha}\notag\\
&&\times
\begin{pmatrix}
1-2f(\omega-\mu_{\alpha})&2f(\omega-\mu_{\alpha})\\
-2f(-(\omega-\mu_{\alpha}))&1-2f(\omega-\mu_{\alpha})
\end{pmatrix}_{kk^{'}}\notag\\ &&
{=}
-i\delta_{jj^{'}}\sum_{\alpha}\Theta_j^\alpha\Gamma_{\alpha}\notag\\
&&\times\left[(2\delta_{kk^{'}}-1)\sgn(\omega-\mu_{\alpha})+{\it k}(\delta_{{\it kk}^{'}}-1)\right]
\label{Sigma_bath}
\end{eqnarray}
incorporates effects of the coupling between the dots and leads, where $\Theta_j^\alpha=\delta_{\alpha L}\delta_{j1}+\delta_{\alpha R}\delta_{j4}$. In the above equations $\Gamma_{L(R)}=\pi|t_{1(4)}^{L(R)}|^{2}\rho_{\text{lead}}$ is an energy independent hybridization strength, where $\rho_{\text{lead}}$ represents the local density of states at the last site of the left or right lead (the leads are equivalent), $f\left(\omega-\mu_{\alpha}\right)=\theta(\mu_\alpha-\omega)
$ is the Fermi-Dirac distribution function of the lead  $\alpha$ with the chemical potential $\mu_{\alpha}$ at zero temperature $T=0$ ($\theta(x)$ is the Heaviside step function).
Throughout this paper we use the notation $k(k^{'})=\pm 1$ for the Keldysh indices $k(k^{'})=\pm$. Finally, the out of equilibrium regime of the system is set by applying the bias voltage $V$ between the leads and choosing $\mu_{L}=-\mu_{R}=V/2$.\par
In order to approximately determine the self-energy $\Sigma$, which accounts for the effects of the electron interaction $U$, and the corresponding Green function $\mathcal{G}$, which is considered to be a matrix ($8 \times 8$ for each spin projection) in the Keldysh-dots space,
we use the the functional renormalization group method in the Keldysh formalism~\cite{Metzner_2012,Gezzi_2007,Jakobs_2007}. This method yields an infinite hierarchy of differential flow equations for the cutoff-parameter $\Lambda$-dependent self-energy $\Sigma^\Lambda$, two-particle and the high-order interaction vertices, which has similar structure 
to the fRG on the Matsubara frequency axis \cite{Karrasch_2006,Metzner_2012}.
In the present study we consider only the flow of the self-energy and neglect the frequency dependence of the self-energy and the flow of the two-particle and higher order vertex functions. It was shown that neglecting frequency dependence of the self-energy allows one to describe both, equilibrium \cite{Karrasch_2006} and non-equilibrium properties \cite{Gezzi_2007,Jakobs_2007}, as well as to access the SFL state \cite{PK_2016,PK_2017}. On the other hand, neglecting flow of two-particle and higher order vertices is sufficient to reproduce the Kondo behavior of the linear conductance of a single quantum dot\cite{Karrasch_2006}; this approach demonstrates an excellent agreement with the density matrix renormalization group (DMRG) and the NRG results for the interacting resonant level model  \cite{Karrasch_2010} and allows us to fulfill exactly charge conservation, which is typically violated in higher order truncations \cite{Gezzi_2007}. 

At the considering level of truncation the above described approximations lead to the closed zero-temperature fRG flow equation for the 
self-energy $\Sigma^{\Lambda}$, which has the form~\cite{Gezzi_2007} 
\begin{equation}
\partial_{\Lambda}\Sigma^{kk^{'};\Lambda}_{jj^{'};\sigma}=-ikU\delta_{kk^{'}}\delta_{jj^{'}}\int{\dfrac{d\omega}{2\pi}\mathcal{S}^{kk;\Lambda}_{jj;\bar{\sigma}}}\left(\omega\right),
\label{fRG_Eq}
\end{equation}
where
$
\mathcal{S}^{kk^{'};\Lambda}_{jj^{'};\sigma}=-\sum_{i i^{'}}\sum_{q q^{'}}\mathcal{G}^{kq^{'};\Lambda}_{ji^{'};\sigma}\partial_{\Lambda}\left[
\Sigma^{\Lambda}_{\rm cut}
\right]^{q^{'}q}_{i^{'}i;\sigma}\mathcal{G}^{qk^{'};\Lambda}_{ij^{'};\sigma}
$
is the single-scale propagator and $\mathcal{G}^{\Lambda}=\left[\mathcal{G}^{-1}-\Sigma^{\Lambda}_{\rm cut}-\Sigma^{\Lambda}\right]^{-1}$ is the $\Lambda$-dependent propagator, where
\begin{eqnarray}
\left[\Sigma^{\Lambda}_{\rm cut}\right]^{kk^{'}}_{jj^{'};\sigma}=&-&i\Lambda\delta_{jj^{'}}\left[(2\delta_{kk^{'}}-1){\sgn(\omega)}\right.\notag\\&+&\left.k(\delta_{kk^{'}}-1)\right]
\end{eqnarray}
introduces the $\Lambda$-dependence of $\mathcal{G}$ through the reservoir cutoff scheme~\cite{Karrasch_2010} (note that here $\Sigma^{\Lambda}_{\rm cut}$ is defined in the contour basis ($k(k^{'})=\pm 1$) instead of the retarded-advanced Keldysh basis ($k(k^{'})\in\{r,a,\rm K\}$). For some quantities in the equilibrium we also compare results to those from the fRG approach considering flow of the vertices (the corresponding fRG equations can be found, e.g., in Refs.~\cite{Gezzi_2007,Jakobs_2007}).\par
By solving the differential equation (\ref{fRG_Eq}) with the initial condition $\Sigma^{\Lambda_{\rm ini}}=0$, where $\Lambda_{\rm ini}$ is some initial scale, which is chosen to be much larger than all energy scales of the quantum dot system (note that we have included the term $U/2$ into the quadratic part of the Hamiltonian in Eq. (\ref{H_dot})), at the scale $\Lambda=0$ we obtain the energy-independent approximation to the self-energy $\Sigma=\Sigma^{\Lambda\rightarrow 0}$. To induce small initial spin splitting, which can be further enhanced by correlation effects during fRG flow (and therefore allows us to obtain local moments), we apply small magnetic field $H/{\rm max}(\Gamma_{L,R})=0.001$. Due to use of truncation (\ref{fRG_Eq}) of fRG hierarchy at first (self-energy) instead or second order (vertices), the counterterm technique, suggested in previous studies~\cite{PK_2017,PK_2016} is not necessary, and does not change the obtained results.
\section{Local moments and conductance in the equilibrium regime $(V=0)$}\par

Let us first consider the results of the application of outlined fRG approach
in the equilibrium $(V=0)$. This case was intensively studied within equilibrium fRG for DQD system (see, e.g., Refs. \cite{PK_2016,PK_2017}), where good agreement with numerical renormalization-group (NRG) results was obtained.
As in previous study of two parallel quantum dots~\cite{PK_2017,PK_2016}, it is convenient to perform transformation of the electronic states on QD2,3 to the even ($e$) and odd ($o$)  orbitals, see Appendix A. In numerical calculations, we set  $\Gamma_{L}=\Gamma_{R}=\Gamma$, $U/\Gamma=2$, $T=0$ and use $\Gamma$ as the energy unit. 

\subsection{Local magnetic moments}

To analyze the presence of the magnetic moment in the system we consider $\epsilon=0$, $t_{14}=0$ case (the results at finite small $\epsilon$ and finite small or moderate $t_{14}$ are qualitatively similar) and calculate the average square of the spin $\langle \mathbf{S}_{e/o}^2 \rangle$, corresponding to the even and odd orbitals, where $\vec{\mathbf{S}}_{p}=({1}/{2})\sum_{\sigma,\sigma^{'}}d^{\dagger}_{p,\sigma}\vec{\bm{\sigma}}_{\sigma\sigma^{'}}d_{p,\sigma^{'}}$ is the spin operator and $\vec{\bm{\sigma}}$ is the vector of the Pauli spin matrices. 

\begin{figure}[t]
\centering
\includegraphics[width=0.8\linewidth]{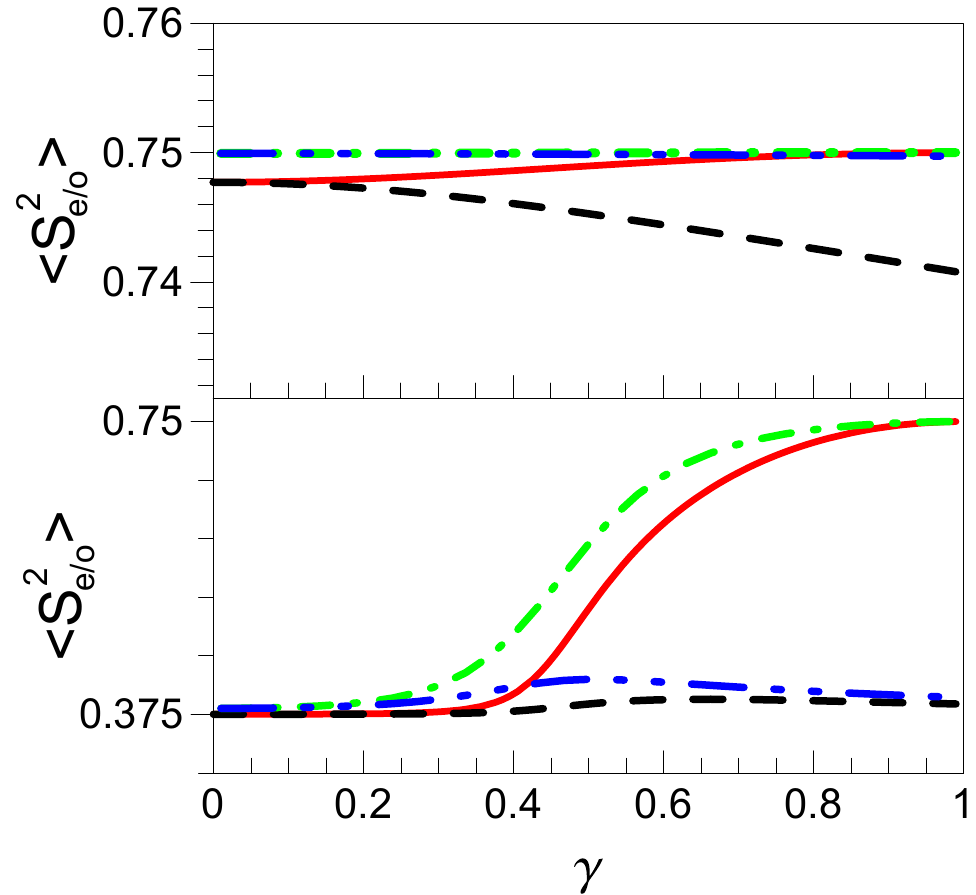}
\caption{(Color online). The average square of a magnetic moment $\langle\mathbf{S}_{e/o}^{2}\rangle$ in the even (dashed black lines) and odd (solid red lines) states as a function of $\gamma$ for $t=0.05$ (upper panel) and $t=0.5$ (lower panel), and $t_{14}=\epsilon=0$. Dashed-dotted-dotted blue and dashed-dotted green lines show  $\langle\mathbf{S}_{e}^{2}\rangle$ and $\langle\mathbf{S}_{o}^{2}\rangle$, respectively, in the fRG approach with the flow of the two-particle vertex (the corresponding curves are almost indistinguishable for $t=0.05$).}
\label{S2even_odd_gamma_b_c}
\end{figure}

Fig.~\ref{S2even_odd_gamma_b_c} shows the dependence of $\langle \mathbf{S}_{e/o}^2 \rangle$ on the parameter $\gamma$ for various values of $t$. 
As one can expect,
for small $t$ (see, e.g.,  $t=0.05$ case) the average $\langle \mathbf{S}_{e/o}^2 \rangle\approx 3/4$, which (together with the filling $\langle n_{e(o),\uparrow}\rangle\approx 1$ and $\langle n_{e(o),\downarrow}\rangle\approx 0$) means that the electron is 
almost localized on both the odd and even orbitals ($\langle n_{e/o,\uparrow}n_{e/o,\downarrow}\rangle\approx 0$) due to weak connection of these orbitals with quantum dots 1,4, namely 
$t_{pq}\ll U$ ($p\in\{1,4\}$, $q\in\{e,o\}$). 
The corresponding 
square of the spins on quantum dots QD2 and QD3,  $\langle \mathbf{S}_{2,3} ^2\rangle\approx 3/4$. 
The average $\langle \mathbf{S}_{o}^2 \rangle$ 
monotonically increases  up to a maximum value of $\langle \mathbf{S}_{o}^2 \rangle=3/4$ at $\gamma=1$ due to 
decrease of the coupling $t_{4o}$ between the odd orbital and quantum dot QD4 (for our definition of the odd orbital $t_{4o}=0$ for $\gamma=1$ and $t_{1o}=0$ for any $\gamma$).
In contrast, both hopping parameters $t_{1e}$ and $t_{4e}$, 
associated with the even orbital, increase with $\gamma$, which leads to a smooth decrease of $\langle \mathbf{S}_{e}^2 \rangle$. 
It is important to note that in this 
and the following cases we find $\langle \mathbf{S}_{1(4)}^2 \rangle$ close to its free-electron value $3/8$, which indicates that there are no local magnetic moments in quantum dots QD1 and QD4.



Increase of the hopping strength $t$ leads to delocalization of the electronic states, which yields a gradual decrease of $\gamma=0$ value of $\langle \mathbf{S}_{e/o}^2 \rangle$. 
Starting with some sufficiently large value of $t$, we find that $\langle \mathbf{S}_{e/o}^2 \rangle\rightarrow 3/8$ for $\gamma\rightarrow 0$, which means that there are no magnetic local moments present in the even/odd states. At the same time, as shown in Fig.~\ref{S2even_odd_gamma_b_c},  with increase of $\gamma$ from $\gamma=0$ to $\gamma=1$, $\langle \mathbf{S}_{o}^2 \rangle$ increases from 
3/8 
to the value 
3/4, showing presence of the local magnetic moment in the odd state at $\gamma\gtrsim 0.6$ (in this case $\langle{n_{o,\uparrow}}\rangle\approx 1$, $\langle{n_{o,\downarrow}}\rangle\approx 0$, $\langle{n_{e,\sigma}}\rangle\approx 0.5$).
This corresponds to the so called singular Fermi liquid state~\cite{Zitko_2007(2),PK_2017,PK_2016} and explained by the fact that, regardless of the choice of $t$, the odd state is almost disconnected from the leads at $\gamma \rightarrow 1$  (in particular, hopping matrix element $t_{4o}$ associated with the odd states, decreases to zero)
and hence, the local magnetic moment on the odd orbital is always well-defined when $\gamma\rightarrow 1$. At the same time, $\langle \mathbf{S}_{e}^2 \rangle\approx 3/8$ remains almost unchanged 
with the variation of $\gamma$, since this orbital remains 
strongly coupled to the quantum dots QD1 and QD4 
, which, in turn, have a direct hybridization with the leads, cf. Ref.~\cite{PK_2017}. 
Thus, in contrast to the cases considered above, in this case only the odd orbital is responsible for the appearance of an unscreened local magnetic moment in the system. 

In order to analyze the role of the neglected vertex corrections, 
we compared the 
obtained results with those from fRG calculations that account for the flow of the two-particle vertex functions, which for DQD system yielded agreement with NRG approach. To eliminate the problem of the divergences of the vertices in the fRG flow, we use the counterterm extension of the fRG approach (related discussion can be found in Refs.~\cite{PK_2016,PK_2017})
with initial magnetic field $\tilde{H}/\Gamma=0.02$, which is switched off linearly with $\Lambda$ starting from the scale $\Lambda_{c}/\Gamma=0.02$. It turns out, that for intermediate and large hopping parameters between the quantum dots ${\rm min}(t_{ij})\gtrsim U,\Gamma$ 
$(i,j\in\{1,2\})$, the renormalization of the two-particle vertex produces only small quantitative changes to the self-energy, obtained from the first-order fRG scheme 
(see, e. g., the results for $\langle \mathbf{S}_{e/o}^2 \rangle$ for $t=0.5$ shown in the lower panels of Fig.~\ref{S2even_odd_gamma_b_c}). In the regime of small hopping strength ${\rm max}(t_{ij})\ll U,\Gamma$ the energy splitting between the spin-up and -down components of the self-energy in the fRG approach with account for the flow of the two-particle vertex is somewhat larger in comparison with that obtained in the first-order fRG approach and account of the flow of the two-particle vertex leads to enhancement of the magnetic moments in the QQD system 
(see upper panel of Fig.~\ref{S2even_odd_gamma_b_c}). However, even in this case, the physical picture of the formation of the magnetic moment in the quantum dot system remains unchanged.

 Note that in the cases when we obtain $\langle {\bf S}^2_{e/o}\rangle\approx 3/4$, the obtained values of the local moments suggest that they are not screened by conduction electrons in the considering case of QQD system (the same applies to DQD system). This can be attributed to presence of the effective hopping between even and odd states via QD4 and strong ferromagnetic correlations between even and odd states, which originate from ferromagnetic correlations between QD2,3 (see Fig. \ref{SiSj_V} below). These ferromagnetic correlations, together with the charge transfer between the leads preclude also the formation of the two-channel Kondo effect (see, e.g., Ref. \cite{TwoChanKondo}). 
We have verified that the same fRG approach for a single quantum dot leads to spin unpolarized solution for $H\rightarrow 0$, which mimics 
screening of the local moment at $T=0$. This approach also describes aspects of Kondo physics, in particular, the Kondo plateau of conductance, as well as it can properly estimate Kondo temperature from the fRG calculation in a finite magnetic field~\cite{Karrasch_2006}. Thus, the considering fRG approach does not lead to an unphysical magnetic solution (even for the first-order truncation of the fRG equations), as it takes place in the mean-field approximation, and hence  in our case the appearance of the (unscreened) local magnetic moment phase at $\gamma$ close to one is not an artifact of the fRG approach.


\subsection{Total conductance}

\begin{figure}[t]
\centering
\includegraphics[width=0.8\linewidth]{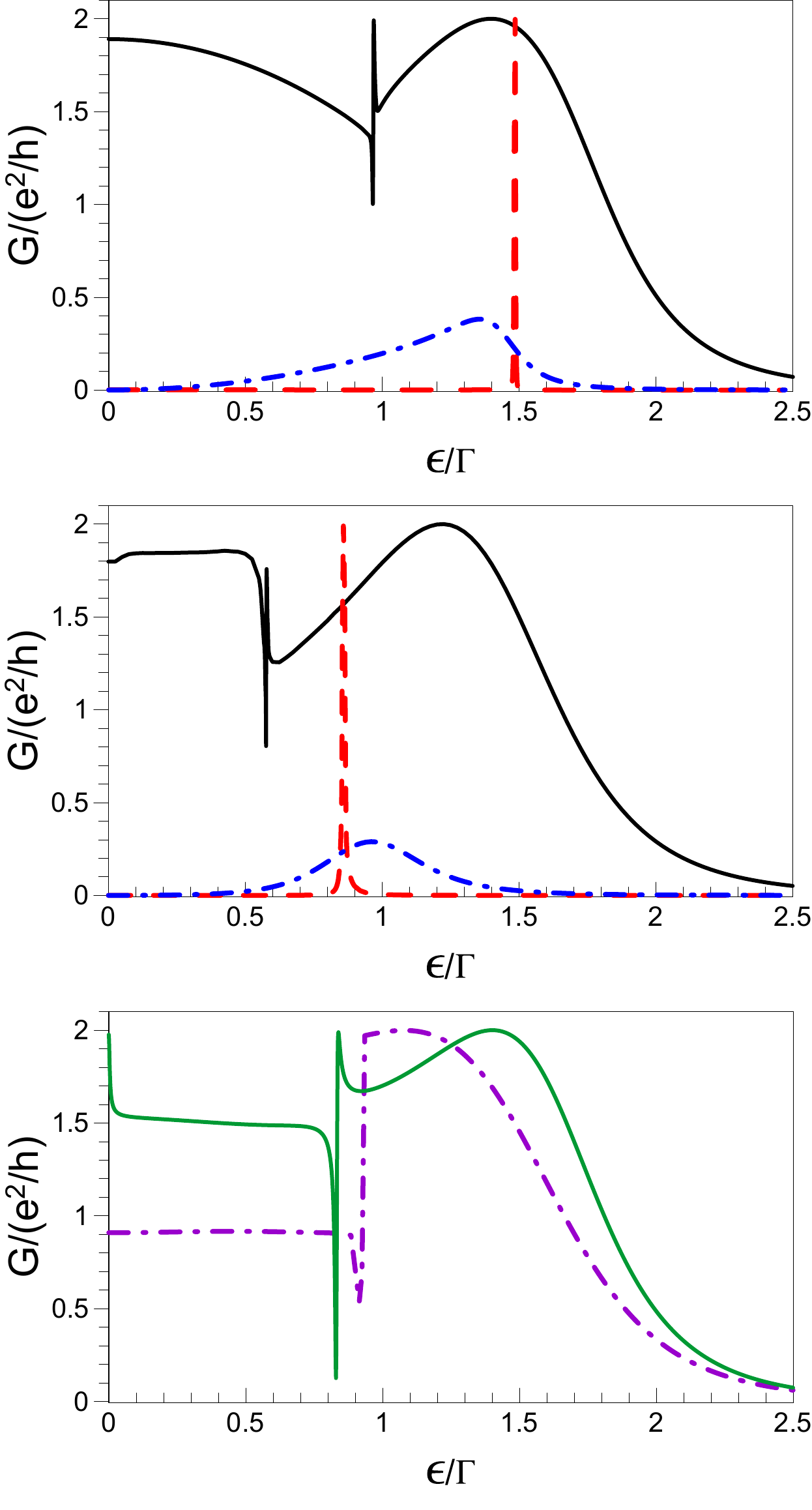}
\caption{(Color online). Upper/middle panel: The gate voltage dependence of the zero-temperature linear conductance $G$ for $(t, \gamma)=(0.05, 0.9)$ (dashed red line), $(t, \gamma)=(0.5, 0.9)$ (solid black line) and $(t, \gamma)=(0.5, 0.1)$ (dashed-dotted blue line) in the fRG approach without/with the flow of the two-particle vertex. 
Lower panel: the conductance in the first-order perturbation theory (solid green line) and in the mean-field approach (dashed-dotted purple line)  for $(t, \gamma)=(0.5, 0.9)$. $t_{14}=0$ for all plots.
}
\label{G_Vg}
\vspace{-0.3cm}
\end{figure}

In Fig.~\ref{G_Vg} 
%
%
%
we present the results for the zero-temperature linear conductance $G=\sum_\sigma G_\sigma$, where $G_\sigma=({4e^{2}}/{h})\Gamma_{L}\Gamma_{R}{\left|\mathcal{G}_{14;\sigma}^{r}\left(\omega=0\right)\right|^{2}}$ as a function of the gate voltage $\epsilon$ for $t_{14}=0$, where $\mathcal{G}^{r}=\mathcal{G}^{--;0}-\mathcal{G}^{-+;0}$ is the retarded Green function in the end of the fRG flow, for various hopping parameters $\left(t,\gamma\right)\in\left\{\left(0.05,0.9\right), \left(0.5,0.9\right), \left(0.5,0.1\right)\right\}$, obtained by numerical integration of 
Eq.~(\ref{fRG_Eq}); the case of fine $t_{14}$ is considered in the next subsection. We use here the Landauer expression for conductance, since we consider $T=0$ case and we have vanishing imaginary part of the self-energy $\Sigma^\Lambda$ in our truncation, which implies physically that we map the interacting system onto the renormalized non-interacting one. 


It can be seen that in the cases $\left(t,\gamma\right)=\left(0.05,0.9\right)$ and $\left(t,\gamma\right)=\left(0.5,0.9\right)$, which
are characterized by the presence of the almost local magnetic moment(s) in the quantum dots at $\epsilon=0$,
the gate voltage dependence of the linear conductance shows abrupt changes 
in the narrow vicinity of some gate voltage. 
This behavior of the conductance is associated with the quantum phase transitions at some critical gate voltage $\epsilon_c$ from the local magnetic moment to the "paramagnetic" regime of the system analogous to the ones which take place in the parallel double dot system~\cite{Zitko_2007(2),PK_2016,PK_2017}. The occupation numbers $\langle n_{e,o}\rangle$ and squares of the local moments $\langle S^2_{e,o} \rangle$ are close to their $\epsilon=0$ values at $|\epsilon|<\epsilon_c$, and correspond to paramagnetic state at $|\epsilon|>\epsilon_c$. 
The dependence of the linear conductance on the gate voltage exhibits near $|\epsilon|=\epsilon_c$ the presence of the asymmetric Fano-like resonance for $\left(t,\gamma\right)=\left(0.5,0.9\right)$, when for $\epsilon=0$
spin-half local magnetic moment is present
and the sharp peak of the conductance for $\left(t,\gamma\right)=\left(0.05,0.9\right)$ case, which in turn corresponds to two spin-half local magnetic moments in the quantum dot ring at zero gate voltage. For the case $\left(t,\gamma\right)=\left(0.5,0.1\right)$, when no magnetic moments exist in the quantum dots, $G\left(\epsilon\right)$ is a smooth nonmonotonic function of $\epsilon$. 
\par

The corresponding results for the linear conductance with account of the vertex flow are presented in the middle panel of Fig.~\ref{G_Vg}.
One can see that the conductance obtained within the 
scheme, which does not include the flow of the two-particle vertex functions, qualitatively reproduces the general patterns and the overall features of the corresponding results with the flow of the vertex. 
It is also necessary to note that, although in all cases the general behavior of the conductance remained the same in the vicinity of the quantum phase transition after accounting for the flow of the two-particle vertex functions, the quantum phase transition point $\epsilon_c$ shifts toward a lower gate voltage.

\begin{figure}[b]
\centering
\includegraphics[width=0.8\linewidth]{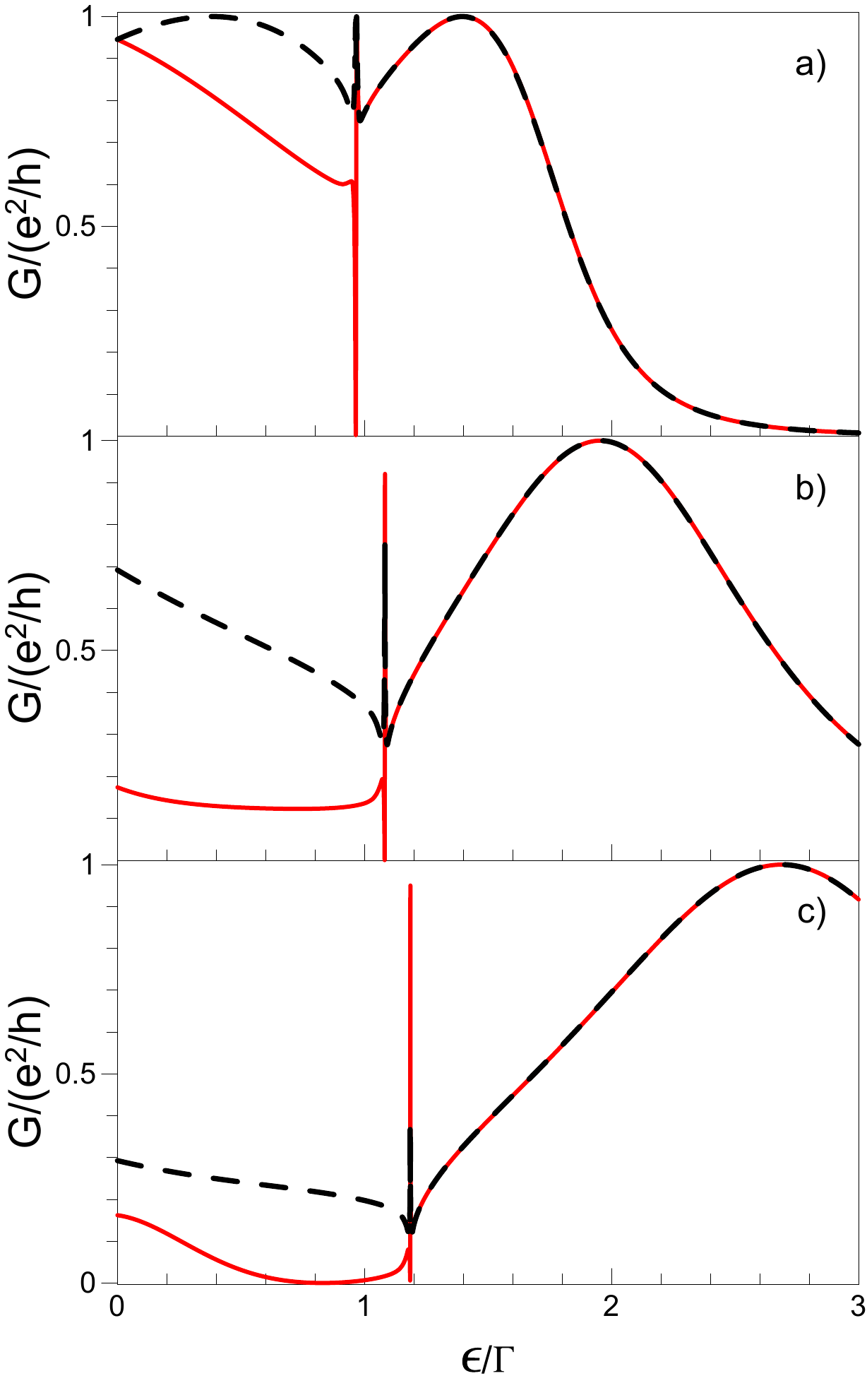}
\caption{(Color online). The gate voltage dependence of the spin-up ($\sigma=\uparrow$, solid red lines) and spin-down ($\sigma=\downarrow$ dashed black lines) zero temperature linear conductance $G_{\sigma}$ for $(t, \gamma)=(0.5, 0.9)$ and $t_{14}=0$ (a), $t_{14}=\Gamma$ (b), $t_{14}=2\Gamma$ (c) in the fRG approach without the flow of the two-particle vertex. }
\label{Gt14}
\end{figure}

\begin{figure}[t]
\centering
\includegraphics[width=0.8\linewidth]{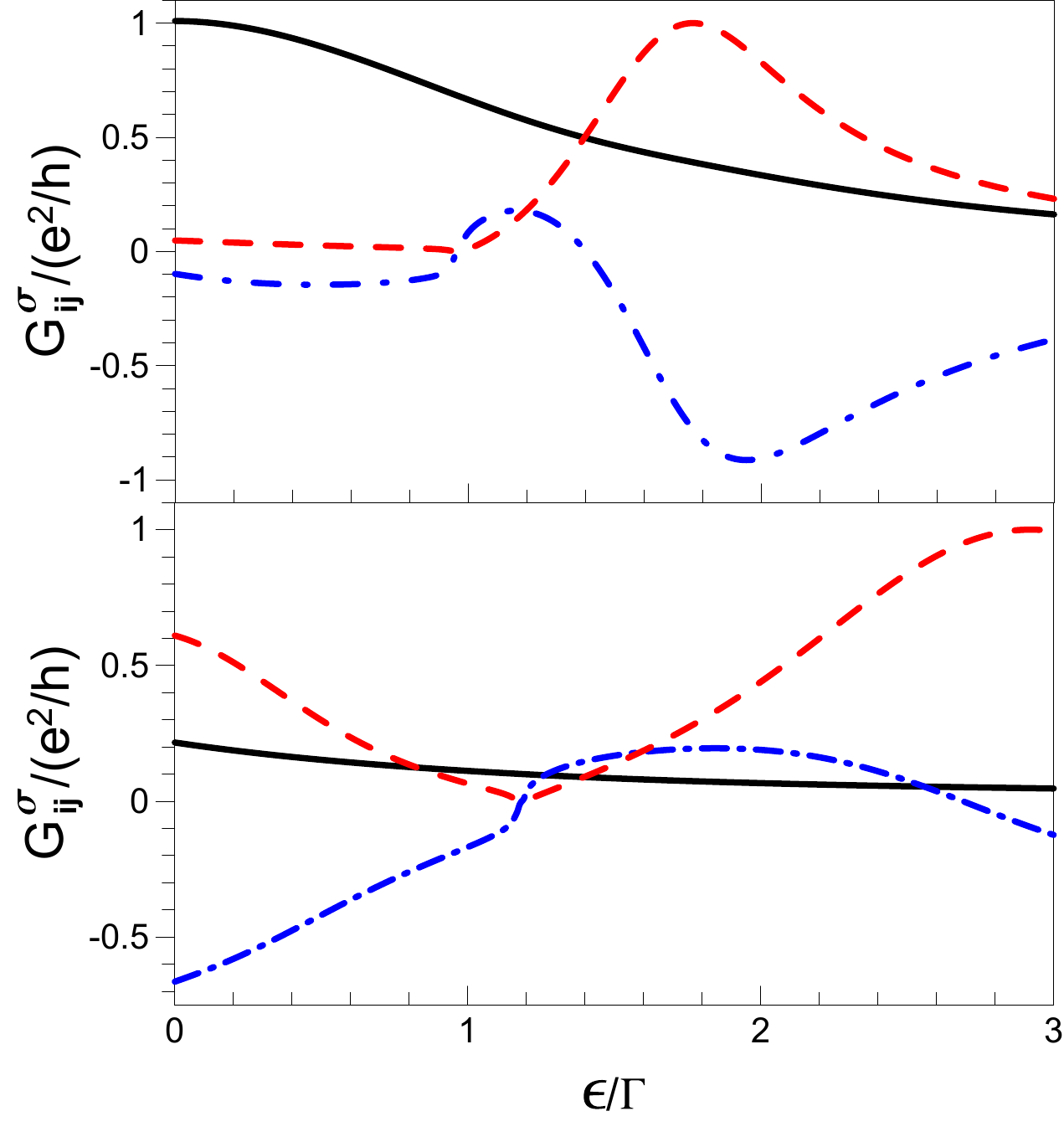}
\caption{(Color online). Conductances $G_{\uparrow,11}$ (solid black line), $G_{\uparrow,{\rm st}}=G_{\uparrow,22}+G_{\uparrow,44}+2G_{\uparrow,24}$ (dashed red line) and the interference contribution $G_{\uparrow,{\rm if}}=2(G_{\uparrow,12}+G_{\uparrow,14})$ (dashed-dotted blue line) as a function of gate voltage $\epsilon$ for $(t,\gamma)=(0.5, 0.9)$, $t_{14}=0$ (upper panel) and $t_{14}=2\Gamma$ (lower panel). 
}
\label{Gij_Vg_t14_0_2}
\end{figure}

To emphasize importance of using fRG approach, which yields non-trivial results already in the truncation, neglecting flow of the two particle vertex, we also show in the lower panel of Fig.~\ref{G_Vg} the results for the linear conductance in the first-order perturbation theory (PT) and mean-field approach (MF) for $\left(t,\gamma\right)=\left(0.5,0.9\right)$. Within the MF approach, the conductance is strongly suppressed near $\epsilon=0$ compared to the fRG results. This is mainly due to the overestimation of the splitting between the spin-up and spin-down states in the MF approach, which does not allow to approach even approximately unitary value of conductance at small $\epsilon$.
At the same time, the MF approach, yielding substantial spin splitting at small  $\epsilon$, is able to predict the existence of the phase with the local magnetic moment. In contrast, the PT theory predicts only the symmetric phase without local moments for all $\epsilon$ 
although the conductance near $\epsilon=0$ in the PT approach is larger than in the MF and somewhat closer to the unitary limit. Note that the resonance near $\epsilon\approx 0.8\Gamma$ in the PT approach is not related to the transition between different magnetic regimes and arises solely due to the interaction-induced dependence of the perturbation theory energy levels of the QQD system on the gate voltage. 

\subsection{Spin-resolved conductances}

In Fig. \ref{Gt14} we consider spin-resolved conductances $G_\sigma(\epsilon)$ in the case of a single local moment $(t, \gamma)=(0.5, 0.9)$ which is most interesting for practical applications, since in this case strong difference between the transport of two spin projections is expected (we still assume vanishingly small magnetic field which orients the local moment along the $z$-axis and therefore creates finite spin splitting of the states, cf. Ref. \cite{PK_2016}). At $t_{14}=0$ we find finite spin-up and spin-down conductances, except the narrow resonance region. While at finite $t_{14}$ the dependence of conductance $G_\downarrow(\epsilon)$ for {minority} spin projection remains qualitatively similar to that for $t_{14}=0$, the conductance for the {majority} spin projection $G_\uparrow(\epsilon)$ is suppressed with respect to $t_{14}=0$ case, and above a certain value of $t_{14}$ vanishes at some gate voltage, forming a plateau with a small, almost vanishing conductance. This vanishing of conductance occurs due to destructive interference of different paths of propagating of spin-up electrons (note that the dependence on the spin occurs due to preferred orientation of the spin of electrons along the field in the even state, which is favored by ferromagnetic correlations between even and odd orbitals and orientation of the local moment along the field).

To get further insight into the mechanism of the conductance in QQD system and its suppression for the majority spin projection,
{we consider partial contributions to the conductance through various states of the system, which energies $\lambda_m$ (including imaginary parts corresponding to the damping due to connection to the leads), $m=1...4$ are determined from the diagonalization of inverse Green function $(\mathcal G^{r}_\sigma(0))^{-1}$ in the end of the flow (due to frequency independence of the self-energy these eigenvalues provide also poles of analytically continued Green function $\mathcal G^{r}_\sigma(\omega)$ in the lower half plane). The obtained eigenstates can be approximately represented as:
\begin{eqnarray}
|\text{es}_1\rangle &\approx& |1\rangle-|4\rangle\notag\\
|{\text{es}}_2\rangle &\approx& \alpha (|2\rangle+|3\rangle)- (|1\rangle+|4\rangle)\notag\\
|\text{es}_3\rangle &\approx& |2\rangle-|3\rangle\notag\\
|{\text{es}}_4\rangle &\approx& \alpha (|1\rangle+|4\rangle)+ (|2\rangle+|3\rangle) \label{QQDstates}
\end{eqnarray}
($\alpha$ depends on the parameters of the system and the spin projection, $|i\rangle$ denotes the state with the considering spin projection $\sigma$ on QD$i$). As it is shown in Appendix B, the states $|\text{es}_{1,2,4}\rangle$ are similar to those in the three quantum dots chain, which corresponds to QD1$\leftrightarrow$(even state of QD2,3)$\leftrightarrow$QD4 subsystem of QQD. In particular, the state $|\text{es}_1\rangle$
describes the resonant tunneling between QD1,4; the states $|{\text{es}}_{2,4}\rangle$ describe sequential tunneling through the even state, as well as the tunneling via the hopping $t_{14}$ (when present). Finally, the state $|\text{es}_3\rangle$ is the odd state of QD2,3, discussed above. By representing $G_{\sigma}=\sum_{mm'}G_{\sigma,mm'}$ where 
\begin{eqnarray}
G_{\sigma,mm' }&=&(4e^2/h)\Gamma_L \Gamma_R {\rm Re} \left[P^{\sigma}_{m} \left(P^{\sigma}_{m'}\right)^*\right],\end{eqnarray}
we individuate the contributions to the conductance through individual eigenstates ($m=m'$) and their interference ($m\ne m'$),
$P^{\sigma}_{m}=U^{\sigma}_{1 m}\left[U^{\sigma}\right]^{-1}_{m 4}/\lambda^{\sigma}_{m},
$
${U}_{im}^{\sigma}$ is the matrix of the eigenvectors of the Green function $(\mathcal G_\sigma^{r}(0))^{-1}$.} We find that the odd state $|\text{es}_3\rangle$ does not contribute to the conductance, except the narrow region of gate voltages near the resonance. 

The other contributions $G_{\uparrow,mm'}$ are shown in Fig.~\ref{Gij_Vg_t14_0_2} where we group together the contributions of states  $|\text{es}_{2,4}\rangle$. 
One can see that for $t_{14}=0$ the biggest contribution to the conductance  comes from the resonant tunneling ($G_{\uparrow,11}$); for $\sigma=\uparrow$ the sequential tunneling contribution $G_{\sigma,\rm st}=G_{\sigma,22}+G_{\sigma,44}+2G_{\sigma,24}$ and its interference $G_{\sigma,\rm if}=2(G_{\sigma,12}+G_{\sigma,14})$ with the resonant tunneling path are small, similarly to the conductance of three-dot chain (see Appendix B). With switching on  hopping $t_{14}$ the situation changes drastically: the resonant tunneling contribution $G_{\uparrow,11}$ is suppressed due to the shift of the energy levels and it becomes comparable to the  contribution from the sequential tunneling $G_{\uparrow,{\rm st}}$. At the same time, these two contributions strongly interfere with each other, such that the total conductance vanishes near $\epsilon=0.8\Gamma$. For another spin projection ($\sigma=\downarrow$, not shown) we find the same resonant tunneling contribution $G_{\downarrow,11}\approx G_{\uparrow,11}$, but for $t_{14}=2\Gamma$ much smaller sequential $G_{\downarrow,\rm st}$ and interference  $G_{\downarrow,{\rm if}}$ contributions than corresponding contributions for $\sigma=\uparrow$; the interference contribution $G_{\downarrow,{\rm if}}$ is also positive for $t_{14}=2\Gamma$. These effects show the possibility of using even a single QQD system as a spin filter in spintronic devices and they can be further enhanced in quantum networks (cf. Ref. \cite{Fu_2012}), which, however, will be studied elsewhere.
We have verified that for DQD system in a similar geometry (with direct hopping between the leads included) the suppression of the majority conductance is much smaller than for QQD system, which is due to absence of resonant tunneling eigenstate $|\text{es}_ 1\rangle$, see Appendix C. 

\section{Non-equilibrium regime $(V\neq 0)$}

\subsection{Local magnetic moments}

Let us consider the impact of non-equilibrium zero-temperature conditions with a finite bias voltage $V$ applied on the local magnetic moments.
We again consider in this subsection the case $t_{14}=\epsilon=0$ (with small finite $\epsilon$ and finite $t_{14}$ yielding qualitatively similar results) and focus on the quantum dot systems with the following hopping parameters $\left(t,\gamma\right)\in\left\{\left(0.05,0.9\right), \left(0.5,0.9\right), \left(0.5,0.1\right)\right\}$, which in the equilibrium case $V=0$ correspond to three different physical situations, discussed in previous subsection: an almost local magnetic moment in both even and odd states (or, equivalently, on the quantum dots QD2 and QD3), the moment in the odd state (i.e. distributed between the QD2 and QD3 quantum dots), and to the absence of a local magnetic moment in the system, respectively.\par

\begin{figure}[b]
\centering
\includegraphics[width=0.8\linewidth]{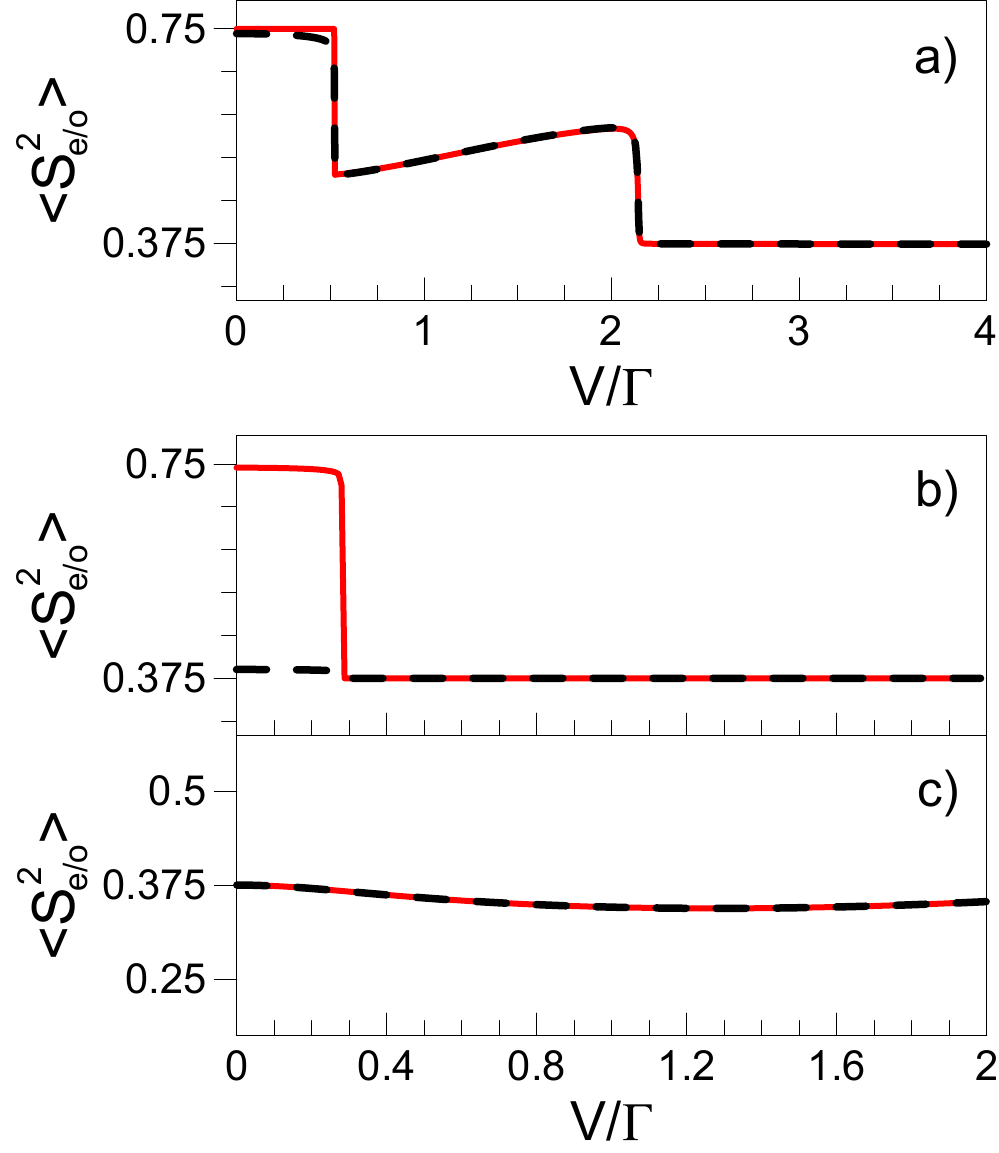}
\caption{(Color online). The average square of a magnetic moment $\langle\mathbf{S}_{e(o)}^{2}\rangle$ in the even (dashed black line) and odd (solid red line) states as a function of bias voltage $V$ for $(t, \gamma)=(0.05, 0.9)$ (a), $(t, \gamma)=(0.5, 0.9)$ (b) and $(t, \gamma)=(0.5, 0.1)$ (c).}
\label{S2_Vn}
\end{figure}

\begin{figure}[t]
\centering
\includegraphics[width=0.8\linewidth]{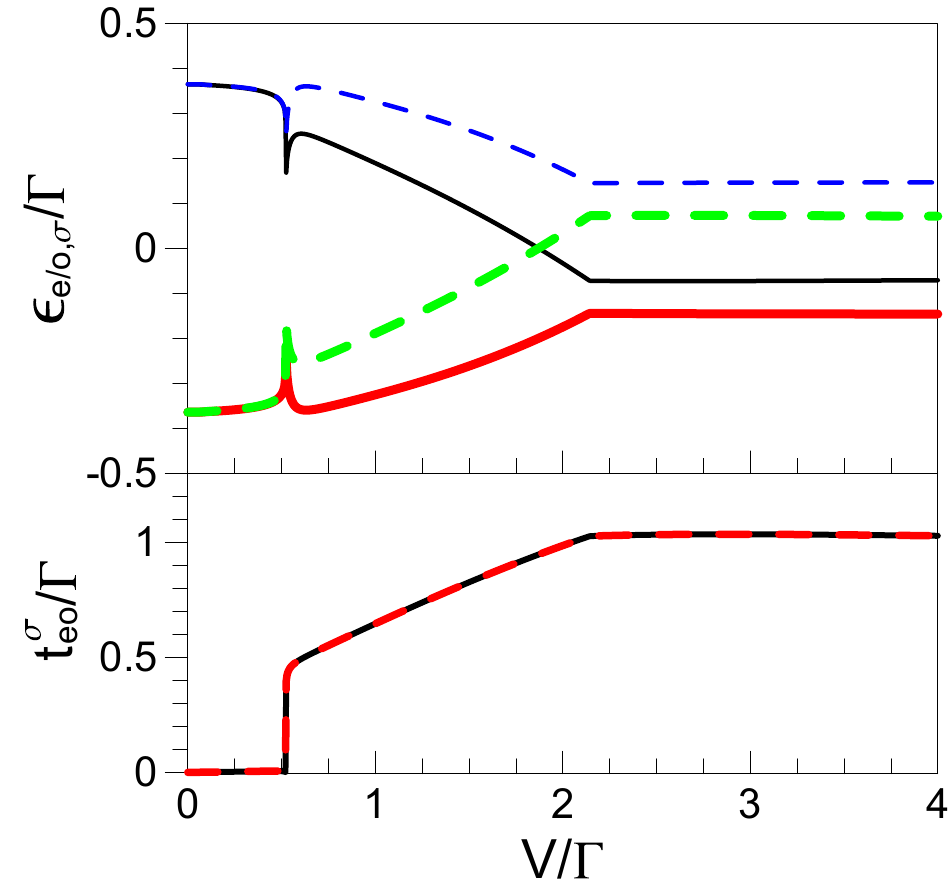}
\caption{(Color online). Upper panel: The renormalized energy levels of the odd states $\epsilon_{o,\sigma}$ (thick solid (red) line for $\sigma=\uparrow$ and thin solid (black) line for $\sigma=\downarrow$) and the even states $\epsilon_{e,\sigma}$ (thick dashed (green) line for $\sigma=\uparrow$ and thin dashed (blue) line for $\sigma=\downarrow$) as a function of bias voltage $V$. 
Lower panel: The renormalized hopping matrix element $t^{\sigma}_{eo}$ (solid black/dashed red line  for $\sigma=\uparrow/\downarrow$) as a function of bias voltage $V$ for $(t, \gamma)=(0.05, 0.9)$.
}
\label{SE_e_o}
\end{figure}

The dependencies of the average square of the spin $\langle \mathbf{S}_{e/o}^2 \rangle$ in the even and odd orbitals on bias voltage $V$ for the first case $\left(t,\gamma\right)=(0.05,0.9)$ are shown in Fig.~\ref{S2_Vn}a.
One can see that increasing bias voltage suppresses the equilibrium value of $\langle \mathbf{S}_{e/o}^2 \rangle$, leading to a double-step behavior, which is related to the strong non-linear change of the renormalized system parameters with the bias voltage. In 
Fig.~\ref{SE_e_o} we plot the renormalized energy levels of the even/odd orbitals $\epsilon_{e/o,\sigma}$ and hopping parameters $t^{\sigma}_{eo}$ (the other system parameters are not renormalized) as a function of $V$. We can see that at not too large $V<0.5\Gamma$
the increase of the bias voltage does not lead to a significant change of the renormalized parameters relative to their equilibrium $(V=0)$ values
and $t^{\sigma}_{eo}$ is pinned to zero as shown in the lower panel of Fig.~\ref{SE_e_o}. 
Therefore, all non-zero hopping parameters are proportional to $t$ and small because of the initial choice of $t$ ($t/\Gamma=0.05$). In this case, the energy levels of the isolated quantum dot system (eigenvalues of the effective non-interacting Hamiltonian) $E_{j,\sigma}$ ($j=1,4$) can be roughly estimated as a set of one-particle energy levels $\{E_{j,\sigma}\}\approx \{\epsilon_{j,\sigma}\}$. This approximation and the observation that within the considered bias voltages range $\epsilon_{e/o,\uparrow}<\mu_{R}=-V/2$ and $\epsilon_{e/o,\downarrow}>\mu_{L}=V/2$
(see Fig. ~\ref{SE_e_o}) allows us to conclude that
$\langle n_{e/o,\uparrow}\rangle \approx 1$ and $\langle n_{e/o,\downarrow}\rangle \approx 0$ for these values of $V$, which reflects formation of local magnetic moment with the spin, aligned along infinitesimally small magnetic field.  

\begin{figure}[t]
\centering
\includegraphics[width=0.8\linewidth]{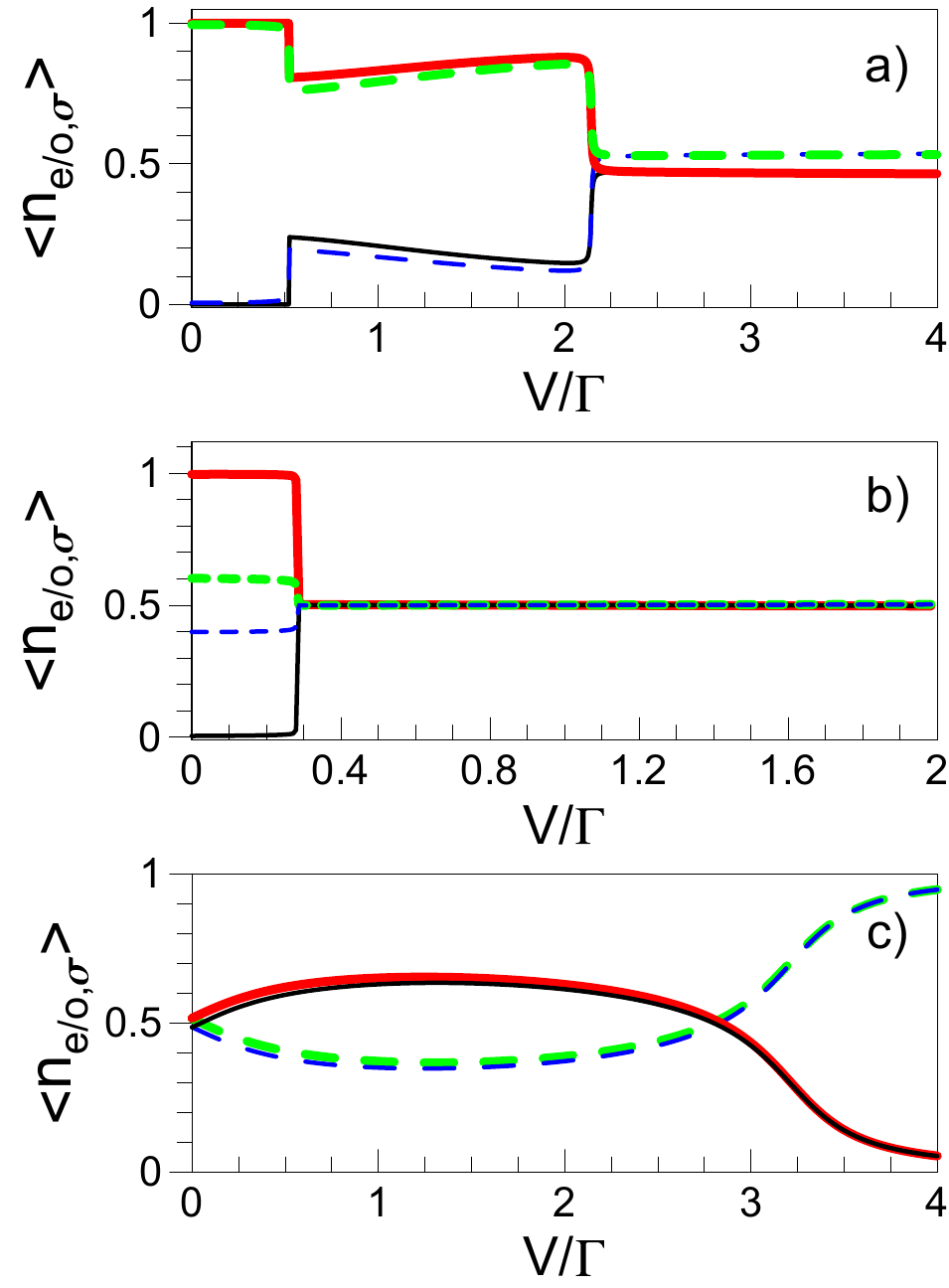}
\caption{(Color online). The occupation numbers in the odd orbitals $\langle n_{o,\sigma}\rangle$ (thick solid (red) line for $\sigma=\uparrow$ and thin solid (black) line for $\sigma=\downarrow$) and the even orbitals $\langle n_{e,\sigma}\rangle$ (thick dashed (green) line for $\sigma=\uparrow$ and thin dashed (blue) line for $\sigma=\downarrow$) as a function of bias voltage $V$ for $(t, \gamma)=(0.05, 0.9)$ (a), $(t, \gamma)=(0.5, 0.9)$ (b) and $(t, \gamma)=(0.5, 0.1)$ (c).}
\label{Neo_V}
\end{figure}
\begin{figure}[t]
\centering
\includegraphics[width=0.8\linewidth]{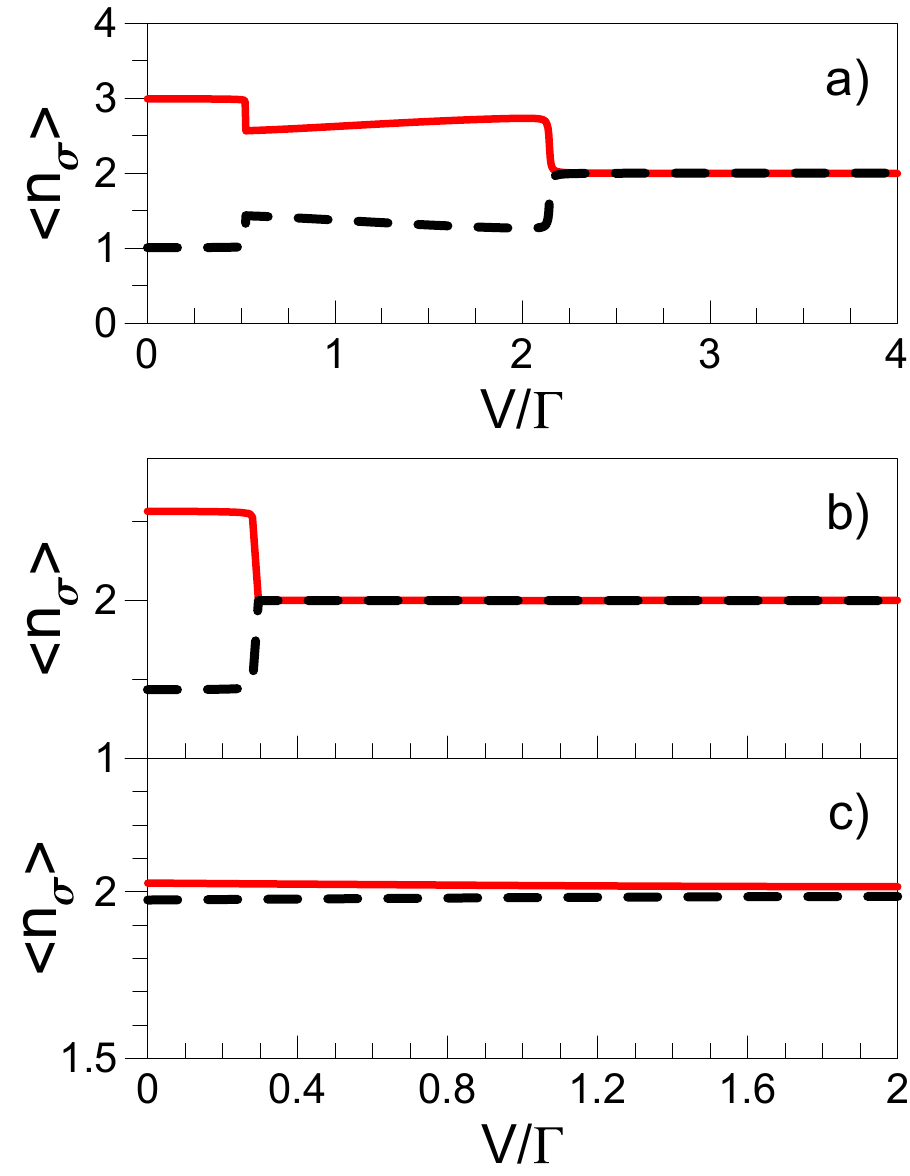}
\caption{(Color online).  The total occupation
number of the spin--up states $\langle n_{\uparrow}\rangle$ (solid red line) and spin--down states $\langle n_{\downarrow}\rangle$ (dashed black line) as a function of bias voltage $V$ for $(t, \gamma)=(0.05, 0.9)$ (a), $(t, \gamma)=(0.5, 0.9)$ (b) and $(t, \gamma)=(0.5, 0.1)$ (c).}
\label{NN_V}
\end{figure}

With further increase of the bias voltage the renormalized energy levels $\epsilon_{e/o,\sigma}$ corresponding to different spin projections  approach each other (see Fig.~\ref{SE_e_o}), and, therefore, the spin splitting 
decreases with $V$. 
It is important that the spin 
splitting does not collapse completely 
even for sufficiently large values of the bias voltage. 
In Fig.~\ref{SE_e_o} we observe the region of the intermediate voltages $0.5 \lesssim V/\Gamma\lesssim 2.1$ for which the splitting of the energy levels is still significant. In contrast to the above considered case, the 
bias voltages in this range lead to the appearance of a nonzero hopping amplitude between the even and odd orbitals $t^{\sigma}_{eo}\gg t$, which increases monotonically with increasing bias voltage, does not depend on the spin orientation and provides additional hybridization of the even/odd states due to the appearance of new possible paths between these states and the leads.
The combined effect of sharp increase of this amplitude and decrease of of the energy levels  splitting, 
results in an abrupt drop of $\langle \mathbf{S}_{e/o}^2 \rangle$ as seen in Fig.~\ref{S2_Vn}a. The values of the square of the moment $\langle \mathbf{S}_{e/o}^2 \rangle$ in the range $0.5 \lesssim V/\Gamma\lesssim 2.1$ 
is different from the non-interacting value $3/8$ due to correlations. This intermediate state can be considered as obeying fractional quasi-local magnetic moment in even and odd orbitals, which appearance is possible entirely due to considered non-equilibrium conditions. 
In the regime of high bias voltage $V \gtrsim 2.1\Gamma$ the even/odd spin-up and -down states are only slightly split and $t^{\sigma}_{eo}$ practically does not change with increasing $V$ (see Fig.~\ref{SE_e_o}a). Such a small splitting in the spin space results in the absence of the magnetic moments in the system and we find $\langle \mathbf{S}_{e/o}^2 \rangle \approx \langle \mathbf{S}_{j}^2 \rangle \approx 3/8$.

The calculation of the average occupation numbers 
confirms 
the results, obtained above (see Figs.~\ref{Neo_V}a and \ref{NN_V}a).
In the range $V\lesssim \Gamma/2$ for the quantum dots QD1 and QD4 we find $\langle n_{1(4),\sigma}\rangle\approx 0.5$ (the corresponding bias voltage dependencies are not presented here). Consequently, we have $\langle n_{\uparrow}\rangle\approx 3$ and $\langle n_{\downarrow}\rangle\approx 1$ for the total occupation number of the states spin $\sigma$ projection $\langle n_{\sigma}\rangle=\sum_{j}{\langle n_{j,\sigma}\rangle}$ $(\langle n_{2,\sigma}\rangle+\langle n_{3,\sigma}\rangle=\langle n_{e,\sigma}\rangle+\langle n_{o,\sigma}\rangle)$, and therefore, $\langle n_{\uparrow}\rangle-\langle n_{\downarrow}\rangle\approx 2$ for these bias voltages (note, that we consider only the half-filling case $\epsilon=0$ and $H\rightarrow 0$, which implies that  $\langle n\rangle=\langle n_{\uparrow}\rangle+\langle n_{\downarrow}\rangle=4$). Thus, one can conclude that at bias voltages $V\lesssim \Gamma/2$ 
the values of the occupation numbers and spin-spin correlation functions 
almost coincide with the equilibrium ones.  
For larger $V$
the obtained occupation numbers $\langle n_{e/o,\uparrow}\rangle$ ($\langle n_{e/o,\downarrow}\rangle$) 
are less (greater) than those for the case of $V\lesssim \Gamma/2$ (see Fig.~\ref{Neo_V}a). However, 
the difference between the occupation numbers of spin-up and spin-down states still remains significant in the range $0.5 \lesssim V/\Gamma\lesssim 2.1$. 
As can be seen from the Fig.~\ref{Neo_V}a, in case $V \gtrsim 2.1\Gamma$ we have $\langle n_{e/o,\uparrow}\rangle \approx\langle n_{e/o,\downarrow}\rangle\approx 0.5$.

\begin{figure}[t]
\centering
\includegraphics[width=0.8\linewidth]{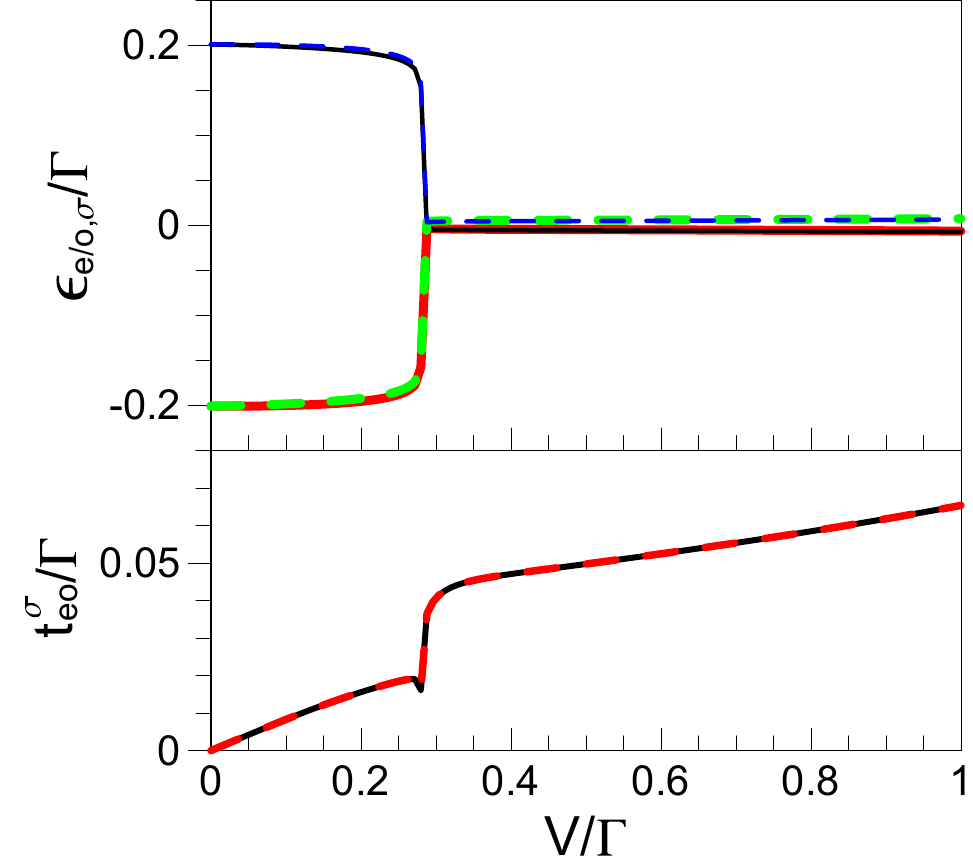}
\caption{(Color online). 
The same as Fig. \ref{SE_e_o} for
$(t, \gamma)=(0.5, 0.9)$.}
\label{SE_e_o_b}
\end{figure}

Let us now consider the case $\left(t,\gamma\right)=(0.5,0.9)$, when the hopping matrix elements $t_{ij}$ are an order of magnitude larger than in the previous case, but have the same ratio between them. In this case the renormalized energy levels $\epsilon_{e/o,\sigma}$ (see Fig.~\ref{SE_e_o_b}) behave near the 
equilibrium
quite analogously to the above considered case $\left(t,\gamma\right)=(0.05,0.9)$, but despite the presence of the large splitting between the spin-up and spin-down states of the even and odd orbitals the appearance of local magnetic moment takes place only on the odd orbital, which is clearly seen from the bias voltage dependence of $\langle \mathbf{S}_{e/o}^2 \rangle$ shown in Fig~\ref{S2_Vn}b. As in the equilibrium case, for $V\lesssim \Gamma/3$ we obtain $\langle \mathbf{S}_{o}^2 \rangle\approx 3/4$, while $\langle \mathbf{S}_{e}^2 \rangle\approx 3/8$.
In contrast to the above considered case of small $t$, the hopping matrix elements $t^{\sigma}_{eo}$ are non-zero even in the low bias region as shown in the lower panel of Fig.~\ref{SE_e_o_b}. However, the generated hopping parameters $t^{\sigma}_{eo}$  are small enough and do not provide 
the hybridization between the odd orbital and the leads sufficient to destroy the magnetic moment. In contrast to the case $(t,\gamma)=(0.05,0.9)$  there is no region of intermediate level splitting, and for $V \gtrsim \Gamma/3$ we have $|\epsilon_{e/o,\uparrow}-\epsilon_{e/o,\downarrow}|\approx H\rightarrow 0$. This leads to the sharp decrease of $\langle \mathbf{S}_{o}^2\rangle$ near the voltage $V=\Gamma/3$ from almost its maximum value of $\langle \mathbf{S}_{o}^2\rangle=3/4$ to $\langle \mathbf{S}_{o}^2\rangle\approx 3/8$ (see Fig.~\ref{S2_Vn}b), such that the magnetic moment is absent for $V \gtrsim \Gamma/3$. As for the above considered case of small $t$, 
we have $\langle n_{o,\uparrow(\downarrow)}\rangle\approx 1(0)$ in the regime with the magnetic moment $(V\lesssim \Gamma/3)$ and $\langle n_{o,\uparrow/\downarrow}\rangle\approx 0.5$ for larger $V$. At the same time, we find that $\langle n_{e,\sigma}\rangle\approx 0.5$ for all values of $V$. As a result, 
the local moment regime is characterized by 
a difference in the total occupation numbers for the spin-up and spin-down states 
approximately equal one 
($\langle n_{\uparrow}\rangle\approx 2.5$ and  $\langle n_{\downarrow}\rangle\approx 1.5$, see Fig.~\ref{NN_V}b). It is worth noting that small difference   between the occupation numbers $\langle n_{e,\uparrow}\rangle$ and $\langle n_{e,\downarrow}\rangle$ for $V\lesssim \Gamma/3$ (see Fig.~\ref{Neo_V}b) is likely due to the overestimation of the spin splitting 
of the energy levels of the even orbital in the fRG scheme, which does not take into account the renormalization of the non-diagonal self-energy elements in the considered order of truncation. This small splitting is not expected to affect the obtained results regarding presence of local magnetic moment at finite $V$.\par

\begin{figure}[t]
\centering
\includegraphics[width=0.8\linewidth]{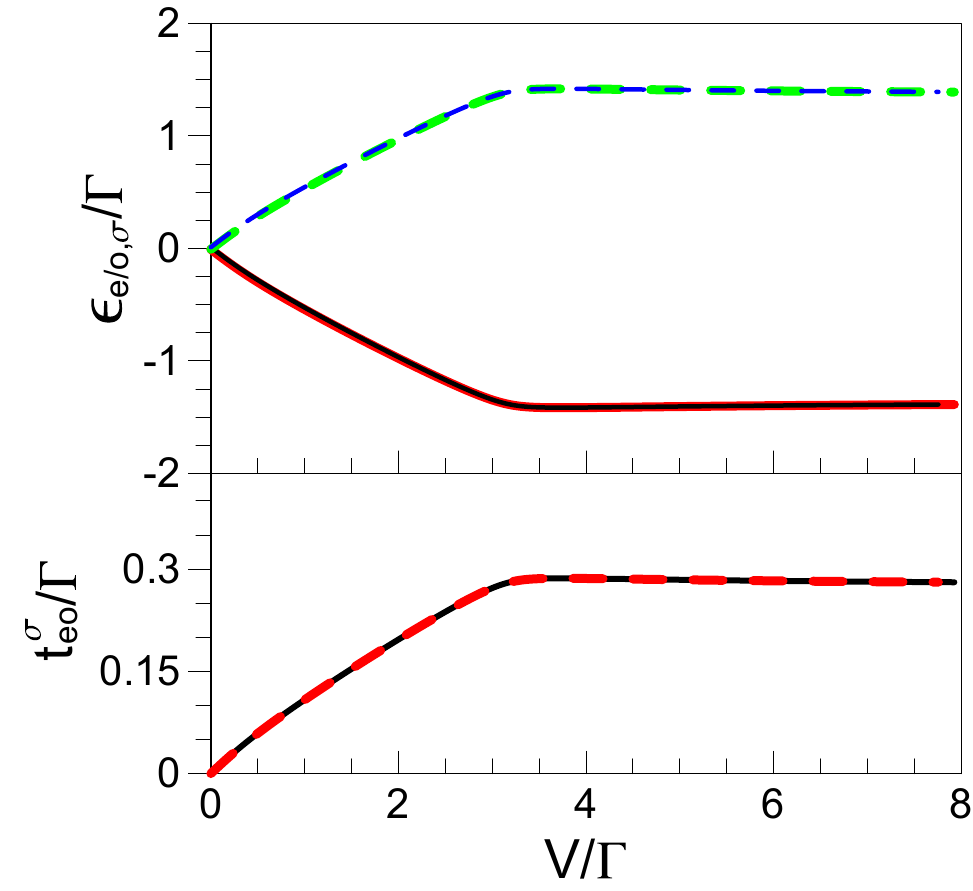}
\caption{(Color online). The same as Fig. \ref{SE_e_o} for
$(t, \gamma)=(0.5, 0.1)$.}
\label{SE_e_o_c}
\vspace{-0.5cm}
\end{figure}

Finally, we consider the case $(t,\gamma)=(0.5,0.1)$ in which the quantum dot system has a strong hopping asymmetry and both the even and odd orbitals are coupled to the quantum dots QD1 and QD4 by almost comparable hopping parameters: $t_{1e}\approx t_{4o}\approx 0.5$, $t_{1o}=0$ and $t_{4e}\approx 0.1$. In this case we do not find any splitting between spin-up and spin-down energy states of the even/odd orbital (see Fig.~\ref{SE_e_o_c}) and, as a consequence, $\langle \mathbf{S}_{e/o}^2 \rangle\approx 3/8$ for an arbitrary bias voltage, as can be seen in Fig.~\ref{S2_Vn}c. In addition, we find a strong renormalization of the energy levels, in particular $\epsilon_{e,\sigma}(\epsilon_{o,\sigma})\propto\mu_{L}(\mu_{R})$ within a wide range of bias voltage near $V=0$ and slowly decreases(increases) with further increase of bias voltage. Note that, $t^{\sigma}_{eo}$ shows linear behavior for bias voltages $V\lesssim 3\Gamma$
and becomes almost constant at higher bias voltages (see the lower panel of Fig.~\ref{SE_e_o_c}).
This behavior of the renormalized parameters leads to the possibility of a significant deviation of the occupations numbers $\langle n_{e/o,\sigma}\rangle$ (see Fig.~\ref{Neo_V}c) from their equilibrium values $\langle n_{e/o,\sigma}\rangle\stackrel{V\rightarrow 0}{\approx} 0.5$, while 
the occupation numbers $\langle n_{\uparrow,\downarrow}\rangle\approx 2$ 
are only slightly different from each other (see Fig.~\ref{NN_V}c). 
In the limit of large bias voltages $V\gg \Gamma$ the occupation numbers converge to $\langle n_{e,\sigma}\rangle=1$
and $\langle n_{o,\sigma}\rangle=0$ 
in contrast to the previous cases, where $\langle n_{e/o,\sigma}\rangle\approx 0.5$ for $V\gg \Gamma$. This behavior
originates from the fact that in the considering case 
the coupling between the even (odd) orbital and the left (right) lead
is much stronger than the corresponding coupling with the 
right (left) lead, which makes the filling of the even (odd) orbital energetically (un)favorable for $V\gg \Gamma$. 
Similar conclusions can be made concerning the fillings at the individual quantum dots, and, as expected from the above qualitative discussion, for $V\gg \Gamma$ we find $\langle n_{1(2),\sigma}\rangle\approx 1$, while $\langle n_{3(4),\sigma}\rangle\approx 0$.\par


\begin{figure}[b]
\centering
\includegraphics[width=0.8\linewidth]{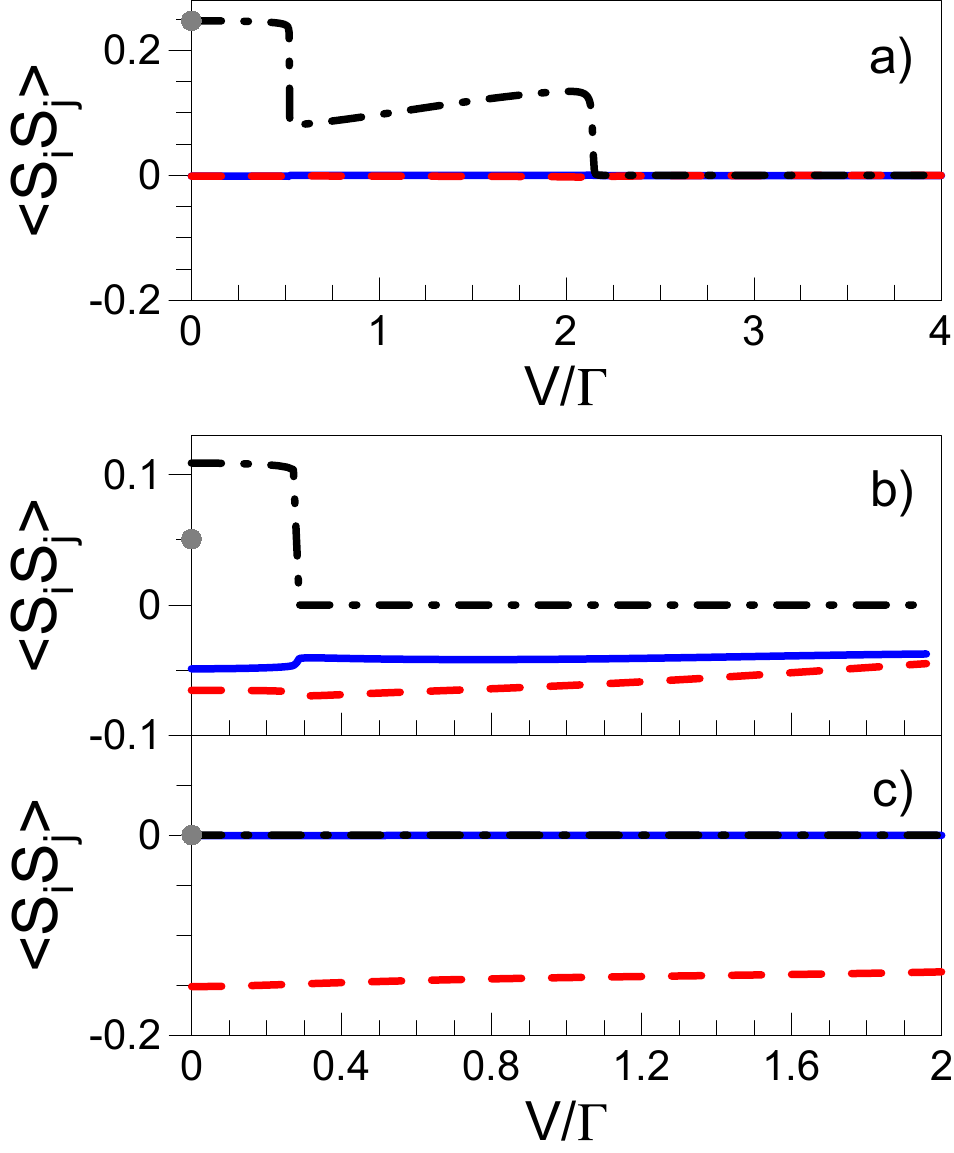}
\caption{(Color online). The spin--spin correlation function $\langle \mathbf{S}_{1}\mathbf{S}_{2}\rangle$ (dashed red line), $\langle \mathbf{S}_{1}\mathbf{S}_{3}\rangle$ (solid blue line) and $\langle \mathbf{S}_{2}\mathbf{S}_{3}\rangle$ (dashed-dotted black line) as a function of bias voltage $V$ for (a) $(t, \gamma)=(0.05, 0.9)$, (b) $(t, \gamma)=(0.5, 0.9)$ and (c) $(t, \gamma)=(0.5, 0.1)$. The value of $\langle \mathbf{S}_{e}\mathbf{S}_{o}\rangle$ at zero bias voltage is indicated by the gray circle.}
\label{SiSj_V}
\end{figure}

The spin-spin correlation functions $\langle \mathbf{S}_{i}\mathbf{S}_{j}\rangle$ corresponding to the above-considered regimes of the system are shown in Figs.~\ref{SiSj_V}a-\ref{SiSj_V}c. One can see that the formation of the magnetic moment in the system is accompanied by the appearance of ferromagnetic correlation between spins on the quantum dots QD2 and QD3, $\langle \mathbf{S}_{2}\mathbf{S}_{3}\rangle>0$, which 
becomes stronger with increasing the value of the magnetic moment. For the regimes
without magnetic moment we find $\langle \mathbf{S}_{2}\mathbf{S}_{3}\rangle\approx 0$. Thus, in the cases $(t, \gamma)=(0.05, 0.9)$ and $(t, \gamma)=(0.5, 0.9)$, $\langle \mathbf{S}_{2}\mathbf{S}_{3}\rangle$ shows step-like behavior as a function of bias voltage. The spin-spin correlation functions $\langle \mathbf{S}_{1}\mathbf{S}_{2}\rangle$=$\langle \mathbf{S}_{3}\mathbf{S}_{4}\rangle$ and $\langle \mathbf{S}_{1}\mathbf{S}_{3}\rangle$=$\langle \mathbf{S}_{2}\mathbf{S}_{4}\rangle$ are always negative (antiferromagnetic) and are proportional in magnitude to the hopping amplitudes between the corresponding quantum dots, i.e, $|\langle \mathbf{S}_{i}\mathbf{S}_{j}\rangle|\sim t_{ij}$. We also note that for all three considered cases the spin-spin correlation between the quantum dots QD1 and QD4 is almost absent, $\langle \mathbf{S}_{1}\mathbf{S}_{4}\rangle\approx 0$.\par

\subsection{Current $J$}

In this subsection we first present zero-temperature results for the $J-V$ characteristics and the bias voltage dependence of the differential conductance $G=e ({dJ}/{dV})$ for the cases considered in the previous subsection. The current with spin $\sigma$ through the lead $\alpha$ is written as~\cite{Meir_1992}
\begin{multline}
J^{\alpha}_{\sigma}=\dfrac{2ie}{h}\Gamma_{\alpha}\sum_{j}{\Theta_j^\alpha}\int{d\omega}\left\{f\left(\omega-\mu_{\alpha}\right)\left[\mathcal{G}_{jj;\sigma}^{r}\left(\omega\right)\right.\right.\\-\left.\left.\mathcal{G}_{jj;\sigma}^{a}\left(\omega\right)\right]+\mathcal{G}_{jj;\sigma}^{-+;0}\left(\omega\right)\right\},
\end{multline}
where  $\mathcal{G}^{a}=\mathcal{G}^{--;0}-\mathcal{G}^{+-;0}$ is the advanced Green function in the end of fRG flow. Using the explicit form of the propagator $\mathcal{G}$ given by Eq.~(\ref{Gf}) we can reduce the above expression to a more convenient form 
\begin{equation}
J^{\alpha=L(R)}_{\sigma}=\dfrac{2ie}{h}{\Gamma_{\alpha}}\sum_j\Theta^\alpha_j\int_{\mu_{r}}^{\mu_{l}}{\mathcal{G}_{jj,\sigma}^{+-(-+);0}\left(\omega\right)d\omega},
\label{Jas}
\end{equation}
where we have used that the non-diagonal components of the self-energy do not flow $\partial_{\Lambda}\Sigma^{kk^{'};\Lambda}_{jj^{'}}\sim\delta_{kk^{'}}\delta_{jj^{'}}$ (see Eq.~\ref{fRG_Eq}) and we have taken the zero-temperature limit for Fermi functions.\par
\begin{figure}[t]
\centering
\includegraphics[width=0.8\linewidth]{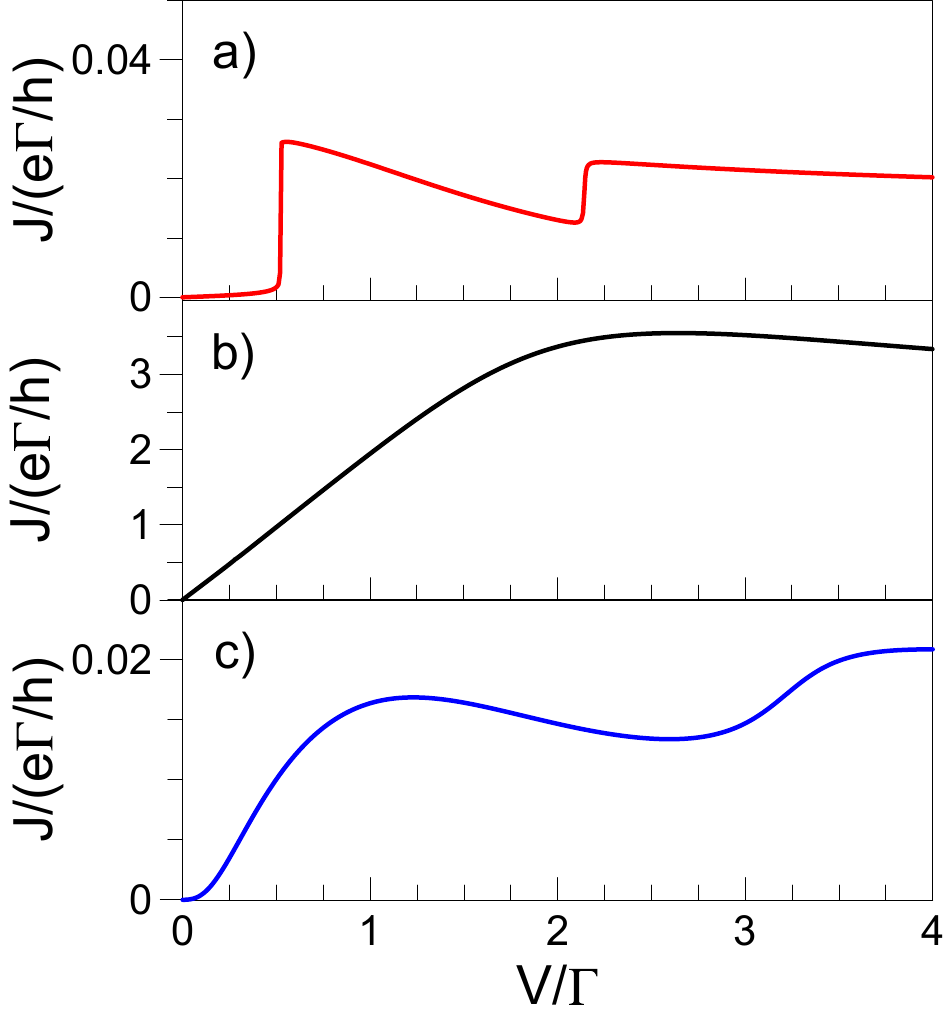}
\caption{(Color online). Zero-temperature current $J$ as a function of bias voltage $V$ for $(t, \gamma)=(0.05, 0.9)$ (a), $(t, \gamma)=(0.5, 0.9)$ (b) and $(t, \gamma)=(0.5, 0.1)$ (c), and $t_{14}=\epsilon=0$.}
\label{J_V1}
\end{figure}
\begin{figure}[t]
\centering
\includegraphics[width=0.8\linewidth]{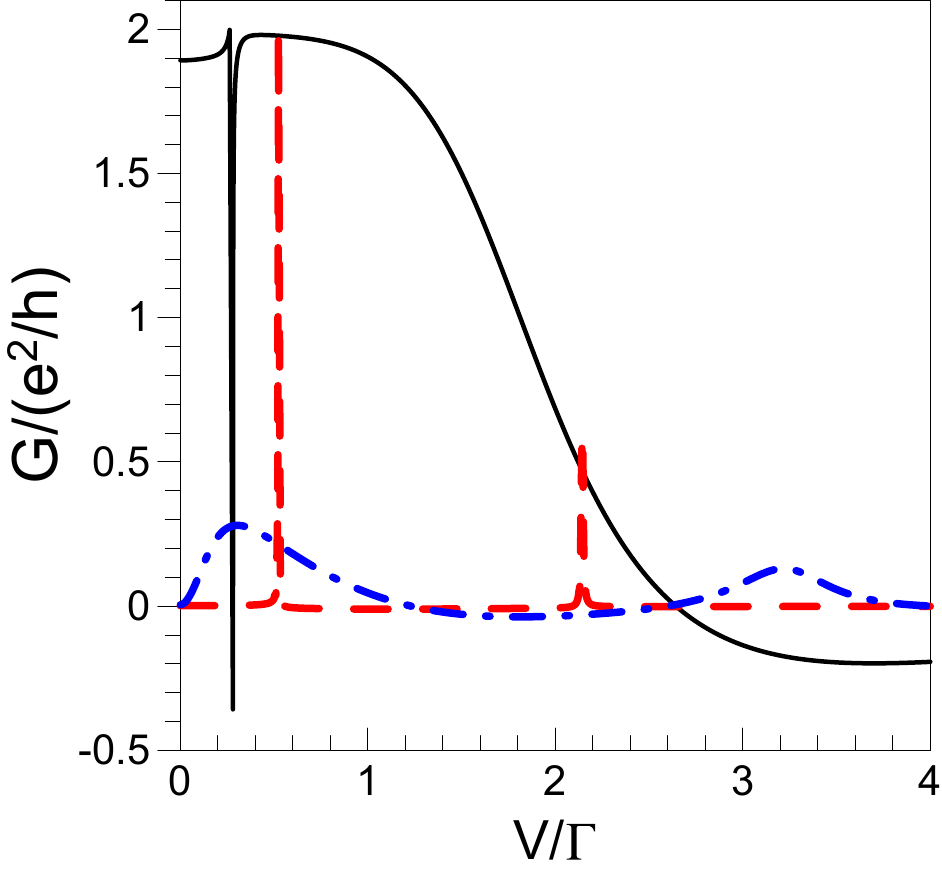}
\caption{(Color online). Zero-temperature differential conductance $G$ as a function of bias voltage $V$ for $(t, \gamma)=(0.05, 0.9)$ (dashed red line), $(t, \gamma)=(0.5, 0.9)$ (solid black line) and $(t, \gamma)=(0.5, 0.1)$ (dashed-dotted blue line: the result for the conductance $G$ was multiplied by 10), and $t_{14}=\epsilon=0$. 
}
\label{G_V}
\end{figure}

The total current $J$ can be calculated as 
\begin{equation}
J=\dfrac{1}{2}\sum_{\sigma}\left(J^{L}_{\sigma}-J^{R}_{\sigma}\right).
\end{equation}
Note that $|J^{R}_{\sigma}|=J^{L}_{\sigma}$ due to the conservation of the current. The dependences of the corresponding currents $J$ 
on the bias voltage $V$ for $t_{14}=\epsilon=0$ are shown in Fig.~\ref{J_V1}. We also plot the zero-temperature differential conductance $G=\sum_\sigma G_\sigma$, where $G_{\sigma}=e ({dJ^{L}_{\sigma}}/{dV})=-e ({dJ^{R}_{\sigma}}/{dV})$, 
in Fig.~\ref{G_V}. 
In the equilibrium limit $V\rightarrow 0$ the current vanishes and for the differential conductance we obtain
\begin{equation}
G_{\sigma}^{0}=\dfrac{ie^{2}}{h}\Gamma_{L}\left[\mathcal{G}_{11;\sigma}^{+-;0}\left(\mu_{L}-0\right)+\mathcal{G}_{11;\sigma}^{+-;0}\left(\mu_{R}+0\right)\right]
\label{G1}
\end{equation}
which coincides with the conductance obtained from the equilibrium Matsubara functional renormalization group method within the Landauer formalism (see Appendix D). In the opposite limit of large bias voltage $V\gg \Gamma$, the current saturates and we find that $G\rightarrow 0$ for all regimes of interest.\par
As one can see from Fig.~\ref{J_V1}a in the case of $(t,\gamma)=(0.05,0.9)$, the $J-V$ curve shows staircase-like structure with two sharp steps, which take place at the same bias voltages, at which  $\langle \mathbf{S}_{e/o}^2 \rangle$ show step-like behavior in Fig.~\ref{S2_Vn}a. As a result, the differential conductance $G$ 
(see Fig.~\ref{G_V}) 
exhibits two narrow peaks located near $V\approx 0.5\Gamma$ and $V\approx 2.1\Gamma$; the first conductance peak almost reaches the unitary limit of the conductance $G=2e^{2}/h$. For bias voltages outside the regions of conductance peaks, we find $G\approx 0$. 
These two peaks are in contrast to the single peak in the gate voltage dependence of the linear conductance at $V=0$ 
(see Fig.~\ref{G_Vg}).
It is also important to note that the $J-V$ characteristic contains regions in which the current decreases with the increase of bias voltage, leading to the negative differential conductance (NDC). As will be shown below, the appearance of NDC is associated with a strong dependence of the renormalized system parameters on the bias voltage, which is in turn induced by the electron-electron interaction.\par
    
For $(t,\gamma)=(0.5,0.9)$, the current shows a small amplitude abrupt jump (not distinguishable in Fig.~\ref{J_V1}b), which is located, as in the above case, at the transition between the different magnetic regimes and results in the asymmetric resonance peak of the differential conductance 
for $V\approx\Gamma/3$. It is interesting to note that the conductance reaches its maximum value in the vicinity of the resonance. Overall in this case, the conductance/current takes significantly higher values compared with those of $(t,\gamma)=(0.05,0.9)$. This holds for $U=0$ and is related to the large coupling strength between quantum dots. In addition,
the conductance becomes negative in two regions of bias voltage: the narrow region near the conductance dip and the semi-infinite one for higher voltages.\par
Finally, in the case of $(t,\gamma)=(0.5,0.1)$, where the magnetic moment is absent for any value of $V$, the current does not show any abrupt behavior and changes smoothly with bias voltage, as shown in Fig.~\ref{J_V1}c. However, the $J-V$ characteristic is strongly non-linear, which
is the result of the non-linear behavior of the renormalized system parameters. 
The NDC effect is also present in this case.\par
As it is evident from the above results, each sharp jump in the current indicates a transition between the regimes with different magnetic moment values. At the same time, a negative differential conductance appears even in the regime without local magnetic moments, as we have shown for $(t, \gamma)=(0.5, 0.1)$ case.  In order to get insight into the origin of the NDC behavior, consider the explicit expression for the zero-temperature conductance $G_{\sigma}$
Direct differentiation of Eq.~(\ref{Jas}) yields $G_{\sigma}=G_{\sigma}^{0
}+G_{\sigma}^{\text{\rm I}}$, where
\begin{equation}
G_{\sigma}^{\text{\rm I}}
=\dfrac{e^{2}}{h}\sum_{p}{K_{p,\sigma}\dfrac{d\epsilon_{p,\sigma}}{dV}}
\end{equation}
with
\begin{equation}
K_{p,\sigma}=2i{\Gamma_{L}}\int_{\mu_{r}}^{\mu_{l}}{\left(\mathcal{G}_{1p;\sigma}^{+-;0}\mathcal{G}_{p1;\sigma}^{--;0}-\mathcal{G}_{1p;\sigma}^{++;0}\mathcal{G}_{p1;\sigma}^{+-;0}\right)d\omega},
\end{equation}
where $p\in\{1,2,3,4\}$.
The contribution $G^{\rm I}$ represents essentially non-equilibrium part of the conductance (which vanishes in the limit $V\rightarrow0$), corresponding to passing the current through each of the quantum dots $p$ and, as shown below it is responsible for the NDC phenomenon (we note that the contributions $G^{\rm 0}_\sigma$ are also affected by finite bias voltage, but remain always positive, see Appendix D).
\begin{figure}[t]
\centering
\includegraphics[width=0.8\linewidth]
{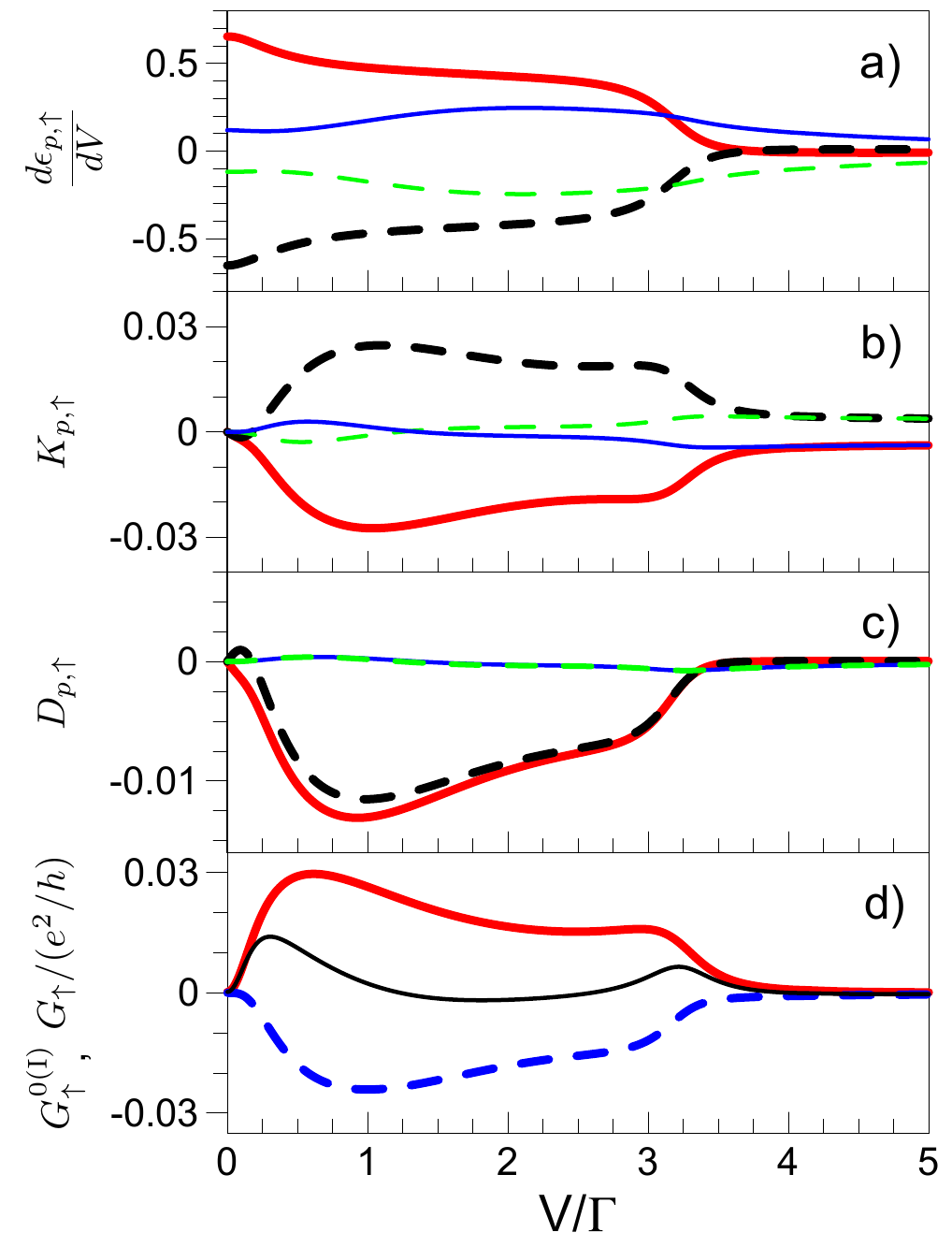}
\caption{(Color online). The panels (a)-(c): The bias voltage dependence of ${d\epsilon_{p,\sigma}}/{dV}$ (a), $K_{p,\sigma}$ (b) and $D_{p,\sigma}$ (c) for $\sigma=\uparrow$. The thin solid (blue), thick solid (red), thick dashed (black), and thin dashed (green) lines correspond to $p=1,2,3$ and 4, respectively. The lower panel (d): The bias voltage dependence of the differential conductance $G_{\sigma}$ (thin solid (black) line), $G_{\sigma}^{0}$ (thick solid (red) line) and $G_{\sigma}^{\text{\rm I}}$ (thick dashed (blue) line) for $\sigma=\uparrow$.}
\label{G12_V}
\end{figure}
\par

As an example, let us analyze the magnitude and sign of the contributions $G_{\sigma}^{0}$ and $G_{\sigma}^{\text{\rm I}}$ to the differential conductance $G_{\sigma}$ for the case $(t,\gamma)=(0.5,0.1)$, $t_{14}=\epsilon=0$, and $\sigma=\uparrow$ (for $\sigma=\downarrow$ we obtain the same results). Note that the conductance $G_{\uparrow}$ reproduces all the features of the total conductance $G$ (see Fig.~\ref{G12_V}c).
The term $G_{\uparrow}^{0}$ is positive for any bias voltage $V$ (see Appendix D), and thus does not contribute to the NDC effect. 

The sign of $G_{\sigma}^{\text{\rm I}}$ is determined by the 
sign of the product $K_{p,\sigma}({d\epsilon_{p,\sigma}}/{dV}).$ 
As shown in Fig.~\ref{G12_V}a, ${d\epsilon_{p,\uparrow}}/{dV}$ can be positive definite $(p=1)$, negative definite $(p=4)$ or even change sign $(p=2,3)$. Moreover, we find that the coefficients $K_{p,\uparrow}$ are also not sign-definite (see Fig.~\ref{G12_V}b). It is important to note that $\left|{d\epsilon_{2(3),\uparrow}}/{dV}\right|>\left|{d\epsilon_{1(4),\uparrow}}/{dV}\right|$ and $|K_{2(3),\uparrow}|\gg|K_{1(4),\uparrow}|$ in a wide region of intermediate values of $V$, which means that terms, corresponding to the contribution of the quantum dots $p=2,3$
give the main contribution to $G_{\uparrow}^{\text{\rm I}}$. This is supported by the bias voltage dependence of the functions $D_{p,\uparrow}=K_{p,\uparrow}({d\epsilon_{p,\uparrow}}/{dV})$ 
shown in Fig.~\ref{G12_V}c. As we can see, $D_{2(3),\uparrow}$ is negative definite (almost everywhere) and have a much greater impact on the conductance, while $D_{1(4),\uparrow}$  is predominantly negative and small in magnitude for all bias voltages. As a result, we find that $G_{\uparrow}^{\text{\rm I}}$ is always negative for arbitrary value of $V$ and is comparable in magnitude with $G_{\uparrow}^{0}$ (see Fig.~\ref{G12_V}c), leading to the strong suppression or even change of sign of the  Landauer-type $G_{\uparrow}^{0}$ contribution to the differential conductance $G_{\uparrow}$. This eventually leads to the appearance of
the NDC effect when the non-equilibrium part dominates, $\left|G_{\uparrow}^{\text{\rm I}}\right|>G_{\uparrow}^{0}$.

Comparing the obtained results for $t_{14}=\epsilon=0$ to those for DQD system (see Appendix C), we find that the double quantum dot system shows a qualitatively similar picture of the magnetic moment(s) and differential conductance as in the QQD system. In particular, as for the QQD system, in the DQD system regimes with two, one or none of the magnetic moment(s) in quantum dots can be realized depends on the choice of the geometry of the system. 
\begin{figure}[t]
\centering
\includegraphics[width=0.8\linewidth]{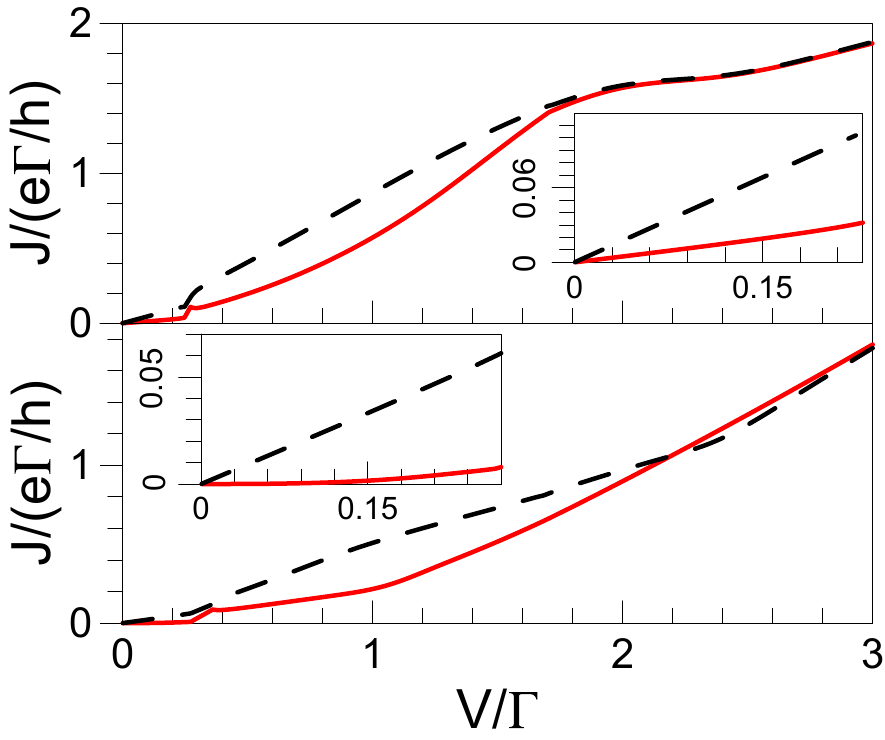}
\caption{(Color online). Zero-temperature current $J_{\sigma}=J^{L}_{\sigma}=|J^{R}_{\sigma}|$ for spin-up ($\sigma=\uparrow$, solid red lines) and spin-down ($\sigma=\downarrow$, dashed black lines) electrons as a function of bias voltage $V$ for $(t, \gamma)=(0.5, 0.9)$, $\epsilon=0.8\Gamma$ and $t_{14}=\Gamma$ (upper panel), $t_{14}=2\Gamma$ (lower panel). Insets zoom the $J_{\sigma}(V)$ dependences at small $V$.}
\label{FigJVt14}
\end{figure}

Finally, we also present the results for the bias voltage dependence of the spin-resolved currents at finite $t_{14}$ (see Fig. \ref{FigJVt14}). In this case we choose $\epsilon=0.8\Gamma$, which corresponds to the gate voltage near the minimum of $G_\uparrow(\varepsilon)$ conductance. One can see that for $t_{14}=2\Gamma$, when in the equilibrium $G_\uparrow(\varepsilon)=0$, the corresponding current $J_\uparrow(V)$ almost vanishes in finite range of bias voltages $V<0.15\Gamma$, and remains small outside this range up to $V\sim\Gamma$. This shows a possibility of spin filtering by QQD even at finite small bias voltages.  

\section{Conclusions}
In summary, in the zero-temperature limit we have discussed the possibility of the formation of the magnetic moments, near equilibrium and the non-equilibrium electron transport in the QQD system coupled to two leads within the non-equilibrium functional renormalization group approach. Our calculations have shown that, depending on the inter-dot coupling (hopping) configuration and bias voltage $V$, different magnetic regimes can be realized in the QQD system.\par 
We have first explored the formation of the magnetic moments in equilibrium $(V=0)$ case. In that case we have shown that the considered fRG approach neglecting vertex flow reproduces qualitatively correct the results, obtained within more sophisticated fRG approach which accounts for the flow of the vertices, and which in turn showed good agreement with the numerical renormalization-group analysis for DQD system. 

We have found three different magnetic regimes that can be achieved in the QQD system by tuning the inter-dot hopping parameters: with two, one or no magnetic moments. As for the parallel double quantum dot system, this difference 
can be understood on the basis of the "even-odd" states. 
The first case (two magnetic moments) corresponds to the situation, where all inter-dot hopping parameters are small compared to the other parameters of the system. 
We have found that the realization of the second and third cases depends on the inter-dot coupling of the "odd" states: well-defined magnetic moment occurs when the coupling of the ”odd” states is sufficiently small. 

While the above mentioned properties are similar to the DQD system, in QQD system the possibility of resonant tunneling between the opposite quantum dots yields somewhat  different transport properties from those in DQD system. This difference becomes especially prominent in the presence of direct hopping between the opposite quantum dots, attached to the leads. In particular, in the presence of this hopping and one local moment in the ring, the conductance of one of the spin projection, oriented along the infinitesimally small magnetic field, is suppressed due to the interference effects, such that the QQD system can be used in spintronic devices.

Then we have considered the influence of the non-equilibrium conditions, appearing because of finite bias voltage, on the above listed magnetic states of the QQD system. We have found magnetic moments (if exist) remain stable in the wide range of voltages near $V=0$. At the same time, for higher bias voltages the destruction of the magnetic state occurs and proceeds in one (two) stage(s) for the QQD systems which coupling configuration allows the formation of the one (two) local moment regime. For the two-stage process the intermediate state possesses fractional magnetic moment.
The current-voltage characteristics and the differential conductances of the system exhibit sharp features at the transition points between different magnetic phases and show negative differential conductance (NDC) behavior. 

It is important to note that although the frequency-independent fRG approximation used in the present study is applicable for the study of the formation of local magnetic moment(s) and transport properties of the quantum dot systems in the regime of small to intermediate Coulomb interactions, it cannot be used to describe spectral functions of the system, as well as various properties associated with the imaginary part of the self-energy, for example, the spin relaxation processes~\cite{Shnirman_2003}. For description of these properties the numerical approaches, in particular the numerical renormalization group, should be farther developed. At the same time, the presented study can help to interpret/achieve new results in experimental realizations of QQD systems, including its use in spintronic devices, as well as to be the guide for studying larger quantum dot and other  and nanoscopic systems, which include closed path (ring) geometries, e.g. organic molecules.\par
{\it Acknowledgements} The work is supported by the theme Quant AAAA-A18-118020190095-4 of FASO Russian Federation, RFBR grant 17-02-00942a and the project 18-2-2-11 of Ural Branch RAS.
\appendix
\section{The transformation to the even and odd orbitals of QD2,3}

Following Refs.~\cite{PK_2017,PK_2016}, it is convenient to perform the canonical transformation from $(\{d_{j,\sigma}\})$ to $(d_{1,\sigma},d_{e,\sigma},d_{o,\sigma},d_{4,\sigma})$ states, where even- $(d_{e,\sigma})$ and odd-parity $(d_{o,\sigma})$ states are defined as
\begin{equation}
\begin{pmatrix}
d_{e,\sigma}\\
d_{o,\sigma}\\
\end{pmatrix}=
\dfrac{1}{\sqrt{1+\eta^{2}}}
\begin{pmatrix}
1 & \eta\\
-\eta & 1\\
\end{pmatrix}
\begin{pmatrix}
d_{2,\sigma}\\
d_{3,\sigma}\\
\end{pmatrix}.
\label{transf}
\end{equation}
Applying the above transformation to the Hamiltonian (\ref{H_dot}) with $U=0$, we get the Hamiltonian of the form 
\begin{eqnarray}
 \mathcal{H}_{\rm QQD}&\stackrel{U=0}{=}&\sum_{p,\sigma}\left(\epsilon_{p}-\sigma H\right)d^{\dagger}_{p,\sigma}d_{p,\sigma}-\sum_{\sigma}{\Bigl[\Bigl(t_{eo}d^{\dagger}_{e,\sigma}d_{o,\sigma}\Bigr.\Bigr.}\notag\\&+&{\left.\left.
t_{1e}d^{\dagger}_{1,\sigma}d_{e,\sigma}+t_{4e}d^{\dagger}_{4,\sigma}d_{e\sigma}\right.\right.}\notag\\&+&{\left.\left.t_{1o}d^{\dagger}_{1,\sigma}d_{o,\sigma}+t_{4o}d^{\dagger}_{4,\sigma}d_{o,\sigma}\right.\right.}\notag\\
&+&{\left.\left.t_{14}d^{\dagger}_{1,\sigma}d_{4,\sigma}\right)\right.}+{\Bigl.\text{H.c.}\Bigr]},
 \label{H_dot_even_odd}
\end{eqnarray}
where $p\in\{1,e,o,4\}$, $\epsilon_{e(o)}=h^{2}\left[\epsilon_{2(3)}+\eta^{2}\epsilon_{3(2)}\right]$ and the non-zero hopping matrix elements between new orbitals are given by
\begin{align}
t_{1e}& = h(t_{12}+\eta t_{13})=h(1+\gamma\eta)t,\notag\\
t_{4e}& = h(t_{24}+\eta t_{34})=h(\gamma+\eta)t,\notag\\
t_{1o}& = h(t_{13}-\eta t_{12})=h(\gamma-\eta)t,\\
t_{4o}& = h(t_{34}-\eta t_{24})=h(1-\gamma\eta)t,\notag\\
t_{eo}& = h^{2}\left[\eta\left(\epsilon_{2}-\epsilon_{3}\right)\right]\notag,
\end{align}
where $h=(1+\eta^{2})^{-1/2}$ and we set $t_{12}=t_{34}=t$, $t_{13}=t_{24}=\gamma t$, where the parameter $\gamma$ varies from zero to unity.\par
\begin{figure}[t]
\centering
\includegraphics[width=0.8\linewidth]{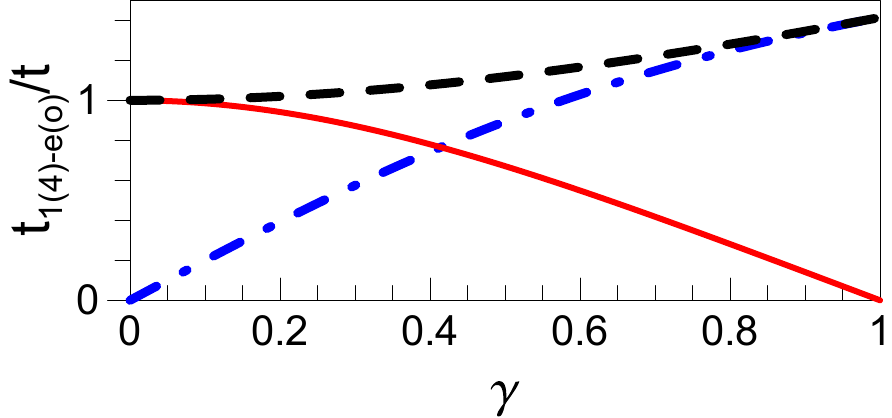}
\caption{(Color online). The hopping matrix elements between the quantum dots QD1/QD4 and the even/odd orbitals $t_{1(4)-e(o)}$ as a function of $\gamma$ for  $\epsilon=0$ (a): $t_{1e}$ (dashed black line), $t_{4e}$ (dashed-dotted blue line), $t_{4o}$ (solid red line).}
\label{S2even_odd_gamma_a}
\end{figure}

We define "odd" orbitals, where the (local) magnetic moment behavior is most pronounced, by choosing the parameter $\eta$ (appearing in Eq.~(\ref{transf}) and has been arbitrary up to this moment) from the condition of the minimum of the coupling strength between the odd orbitals and the quantum dots QD1 and QD4, which is equivalent to finding the minimum of $F(\eta)=|t_{1o}|+|t_{4o}|$. This function has a minimum value of $F=(1-\gamma^{2})(1+\gamma^{2})^{-1/2}t$ for two values of $\eta$, which are $\eta_{1}=\gamma$ and $\eta_{2}=\gamma^{-1}$. It is easy to see that both values of $\eta$ correspond to the same physical situation, and we set $\eta=\gamma$ in the following discussion. Thus, we obtain $t_{1e}=(1+\gamma^{2})^{1/2}t$, $t_{4e}=2\gamma(1+\gamma^{2})^{-1/2}t$, $t_{1o}=0$ and $t_{4o}=(1-\gamma^{2})(1+\gamma^{2})^{-1/2}t$ for the corresponding hopping amplitudes between QD1/QD4 and the even/odd orbitals. Figure~\ref{S2even_odd_gamma_a} displays the ratio $t_{pq}/t$ ($p\in{1,4}$, $q\in{e,o}$) as a function of the "asymmetry parameter" $\gamma$ for $\epsilon=0$. For $\gamma=0$, we have $t_{1e}=t_{4o}=t$ and $t_{4e}=t_{1o}=0$. This result is expected since in that case {quantum dots are connected by $t_{14}$ only (if it is present)}
and the bases coincide: $d_{e,\sigma}\equiv d_{2,\sigma}$, $d_{o,\sigma}\equiv d_{3,\sigma}$. One can see that with increasing symmetry of the quantum dot system (increasing of the parameter $\gamma$), hopping parameters associated with the even-parity states ($t_{1e}$, $t_{4e}$) increase. In contrast, the parameter $t_{4o}$ decreases with $\gamma$ (note, that $t_{1o}=0$). The latter means that the odd-parity orbitals, which are chosen to be the orbitals with the minimal couplings to the others ones and indicate or are responsible for the formation of the magnetic moment in the system, are better and better defined as the symmetry of the hopping parameters between the quantum dots increases. When the system is completely symmetric $(\gamma=1)$, i.e $t_{ij}=t$, we obtain $t_{1e}=t_{4e}=\sqrt{2}t$ and $t_{1o}=t_{4o}=0$, consequently, the odd states are completely disconnected from the rest of the system.\par
On the next step we take into account the on-site electron-electron interaction $U$ through the self-energy obtained from integration of the fRG flow equation~(\ref{fRG_Eq}). It is important to note that from the explicit form of Eq.~(\ref{fRG_Eq}) and the fact that the self-energy $\Sigma$ is frequency independent (and also real) it follows that the Coulomb interaction $U$ does not change the hopping amplitudes $t_{ij}$, ($i\neq j$; $i,j\in\{1...4\}$), and hence $t_{pq}$, ($p\in\{1,4\}$, $q\in\{e,o\}$) at our approximation level. Therefore, in the fRG scheme with the flow of the self-energy only, the initial interacting quantum dot system, {described by Eq.~(\ref{Hamiltonian}) of the main text}, can be considered as the noninteracting one, where energy levels are replaced by renormalized ones: $\epsilon_{j}\rightarrow\epsilon_{j,\sigma}=\epsilon_{j}+\Sigma^{--;\Lambda\rightarrow 0}_{jj;\sigma}$. In the even-odd basis, the parameter $t_{eo}$ is the only hopping term, which can be renormalized by the interaction: $t_{eo}\rightarrow t^{\sigma}_{eo} = h^{2}\left[\eta\left(\epsilon_{2,\sigma}-\epsilon_{3,\sigma}\right)\right]$. Note, that if we take into account of the flow of higher-order vertices (for example, the flow of the two-particle vertex) the situation described above changes, in particular, new hopping matrix elements will be generated. However, as we discuss in the main text of paper these processes do not significantly influence the behavior of local magnetic moments. 


\section{The chain of tree quantum dots attached to the leads}
To explain physical content of the states (\ref{QQDstates}) of QQD system, in this appendix we consider a simpler problem of a three quantum dots chain, which is tractable analytically and models a subsystem QD1$\leftrightarrow|2\rangle+|3\rangle\leftrightarrow$QD4 of QQD system, attached to the leads, where $|2\rangle+|3\rangle$ corresponds to the even energy level of QD2,3 and we assume that the odd energy level of QD2,3 is almost detached from the leads. 

After projecting out the leads the corresponding inverse zero energy retarded Green function of the chain of three quantum dots for each spin projection reads 
\begin{equation}
(\mathcal G^r(0))^{-1}=-\left( 
\begin{array}{ccc}
-i\Gamma  & t & t_{LR} \\ 
t & \epsilon  & t \\ 
t_{LR} & t & -i\Gamma 
\end{array}%
\right) 
\end{equation}
where $\Gamma$ is the hybridization of left and right dots to the leads, $t$ is the hopping between left and middle and between middle and right dots, $t_{LR}$ is the hopping between left and right dots (in this Appendix we denote by $L$ left dot, $M$ -- middle, and $R$ corresponds to the right dot), $\epsilon$ is the energy shift of the middle quantum dot with respect to the left- and right dots. We assume that all effects of the interaction are in the renormalization of these parameters, similarly to the consideration of QQD system in the main text. 

Let us consider first $t_{LR}=0$. Performing the inversion of the Green function, we find the conductance of the chain (per spin projection)
\begin{equation}
G_\sigma=\frac{4e^2}{h}\Gamma^2 |\mathcal G^r_{LR}(0)|^2=\frac{e^2}{h} \frac{1}{1+ \epsilon^2 \Gamma^2/(4t^4)}.
\end{equation}
At $\epsilon$=0 the conductance is unitary due to resonant tunneling of the electrons through the quantum dots. To figure out which states contribute to the conductance we diagonalize the Green function by representing $\mathcal G_{LR}(0)=-\sum_m P_m$
where $P_{m}=U^{\sigma}_{L m}\left[U^{\sigma}\right]^{-1}_{m R}/\lambda^{\sigma}_{m}$ is partial contribution of $m$-th eigenstate, $\lambda_m$ are corresponding eigenvalues and $U_{im}$ are eigenvectors of $-\mathcal G^r(0)^{-1}$. For the considering chain we find the following eigenvectors:
\begin{eqnarray}
|\text{es}_1\rangle &=& |L\rangle-|R\rangle\notag\\
|{\text{es}}_2\rangle &=& 2\alpha |M\rangle- (|L\rangle+|R\rangle)\notag\\
|{\text{es}}_3\rangle &=&|M\rangle+ \alpha  (|L\rangle+|R\rangle)\label{3chainstates}
\end{eqnarray}
where 
\begin{equation}
\alpha=\frac{\sqrt{8 t^2+(\epsilon +i \Gamma)^2}-(\epsilon+i \Gamma)}{4 t}.
\end{equation}
The states (\ref{3chainstates}) are analogous to the states of QQD system (\ref{QQDstates}), except the odd energy level, which is not present in the considering chain. The corresponding eigenvalues
$\lambda_1=-i \Gamma$, $\lambda_{2,3}=(\epsilon-i\Gamma\mp \sqrt{(\epsilon+i \Gamma)^2+8t^2})/2$. Using these values and matrices $U$, we find partial contributions of different states
\begin{eqnarray}
P_1&=&\frac{1}{2i\Gamma}\notag\\
P_{2,3}&=&\mp \frac{4 t^2+\epsilon  \left[ \epsilon+i \Gamma \pm\sqrt{8 t^2+(\epsilon +i \Gamma
   )^2} \right]}{4 \left(2 t^2+i \Gamma  \epsilon\right) \sqrt{8 t^2+(\epsilon +i \Gamma )^2}}
\end{eqnarray}
One can see that at small $\epsilon$ the contributions $P_{2,3}$ almost compensate each other, such that the state $|\text{es}_1\rangle$ is responsible for the resonant tunneling. In the presence of direct hopping $t_{LR}$ between the left- and right- dots we find the corresponding leading contribution to the Green function $P_1=1/(2(i\Gamma+t_{LR}))$, such that the resonant tunneling is suppressed by $t_{LR}$. 

On the other hand, for $\epsilon\gg t,t_{LR},\Gamma$ we find
\begin{eqnarray}
P_2&\simeq-&\frac{1}{2(i
   \Gamma-t_{LR}) }+\frac{t^2}{(i\Gamma-t_{LR}) ^2 \epsilon}\\
P_3&\simeq&\frac{t^2}{\epsilon  ^3}
\end{eqnarray}
such that for small $t_{LR}$ large contribution of the $|\text{es}_1\rangle$ is almost compensated by one for $|\text{es}_2\rangle$, yielding small conductance in the sequential tunneling regime
$G_\sigma\simeq({4e^2}/{h}) {t^4}/({\epsilon^2 \Gamma^2})$.
The contribution of the state $|\text{es}_3\rangle$ is negligible in this case.

\section{Comparison to the double quantum dot system }

\begin{figure}[b]
\centering
\includegraphics[width=0.8\linewidth]{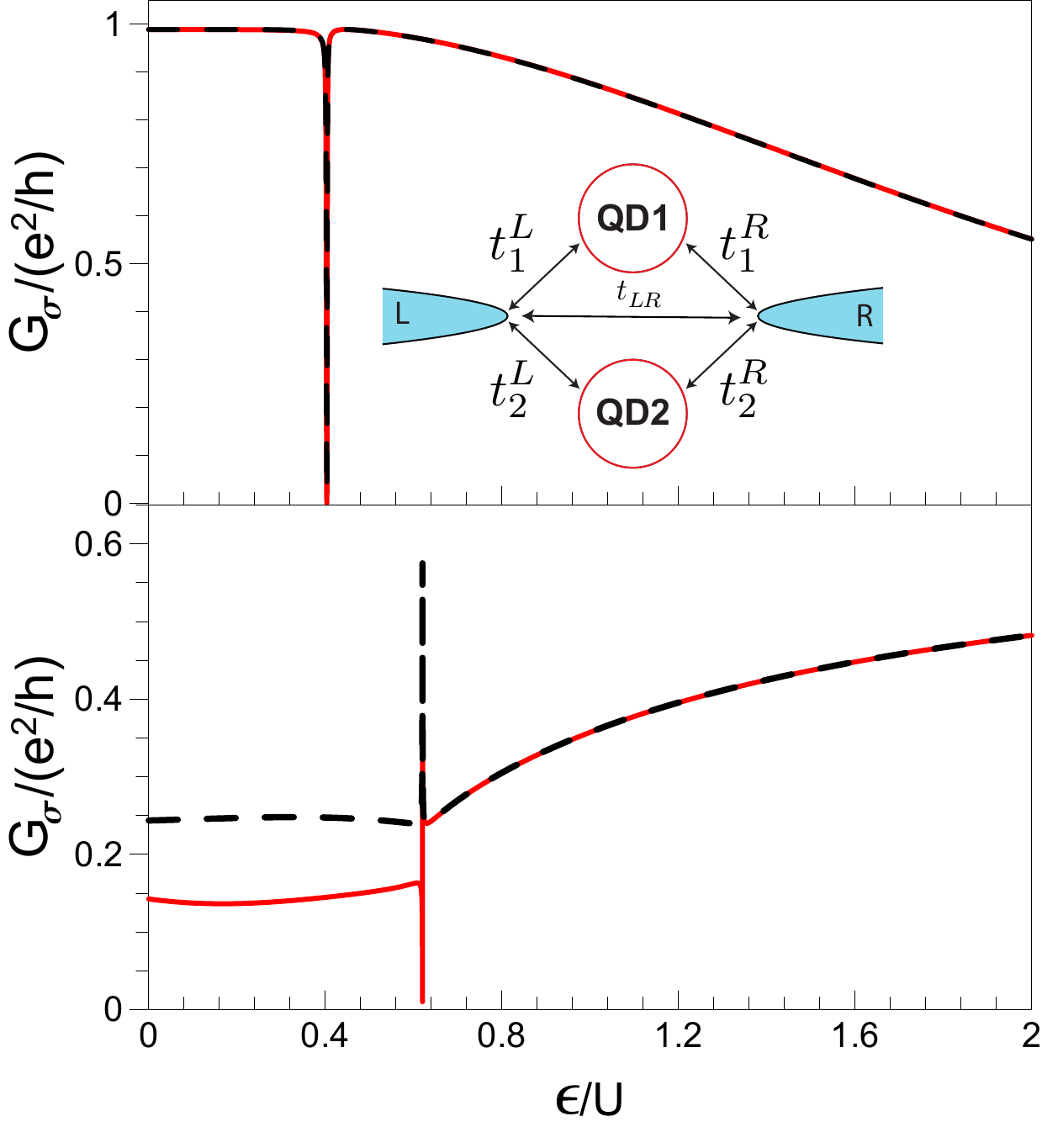}
\caption{(Color online). The gate voltage dependence of the spin-up ($\sigma=\uparrow$, solid red lines) and spin-down ($\sigma=\downarrow$ dashed black lines) zero temperature linear conductance $G_{\sigma}$ of DQD system with  $\gamma=0.9$ with the direct hopping between the left and right lead $t_{LR}=0$ (upper panel) and $(t_{LR}=4 \Gamma)$ (lower panel) in the fRG approach without the flow of the two-particle vertex. Inset: schematic representation of parallel double quantum dots (QD1 and QD2) connected to two left (L) and right (R) leads.}
\label{DQD_G_Vg_up_down}
\end{figure}

In this appendix 
we give a brief analysis of the transport and magnetic properties of the parallel double quantum dot system (DQD), which is schematically shown in Fig.~\ref{DQD_G_Vg_up_down}.

As previously, we focus on the zero-temperature limit. To establish the relationship with the quadruple quantum dot system (QQD) system, we restrict our attention to the case of hopping diagonal asymmetry: $t_{1(2)}^{L(R)}=t$, $t_{1(2)}^{R(L)}=\gamma t$. The hybridization functions $\Gamma_{j}^{\alpha}=\pi|t_{j}^{\alpha}|^{2}\rho_{\text{lead}}$ ($j\in\{1,2\}$, $\alpha\in\{L,R\}$) can be written as $\Gamma_{1(2)}^{L(R)}=\Gamma$, $\Gamma_{1(2)}^{R(L)}=\gamma^{2} \Gamma$. For the both quantum dots we assume equal local Coulomb interactions $U_1=U_2=U$ and equal energy levels $\epsilon_{1,\sigma}=\epsilon_{2,\sigma}=-\sigma H$, with the magnetic field $H/U=0.001$. 

In Fig. \ref{DQD_G_Vg_up_down} we show the gate voltage dependence of the differential conductance for each spin projection in the equilibrium ($V=0$) for $U=2\Gamma$, $t_{LR}=0$ and $t_{LR}=4\Gamma$. For $t_{LR}=0$ we find almost equal conductances of the two spin projections, which is due to rather small spin splitting of the energy levels. The plateau of the conductance at small $\epsilon$ corresponds to the presence of local magnetic moment in the odd orbital and appears due to pinning of the even energy levels to their value at $\epsilon=0$. This pinning and plateau of $G(\epsilon)$ are similar to the Kondo plateau for a single quantum dot \cite{Karrasch_2006}. For rather large $t_{LR}=4\Gamma$ the difference of the conductances of spin-up and spin-down electrons remains small, which is due to the weakness of the interference effects in this case.

\begin{figure}[t]
\centering
\includegraphics[width=0.9\linewidth]{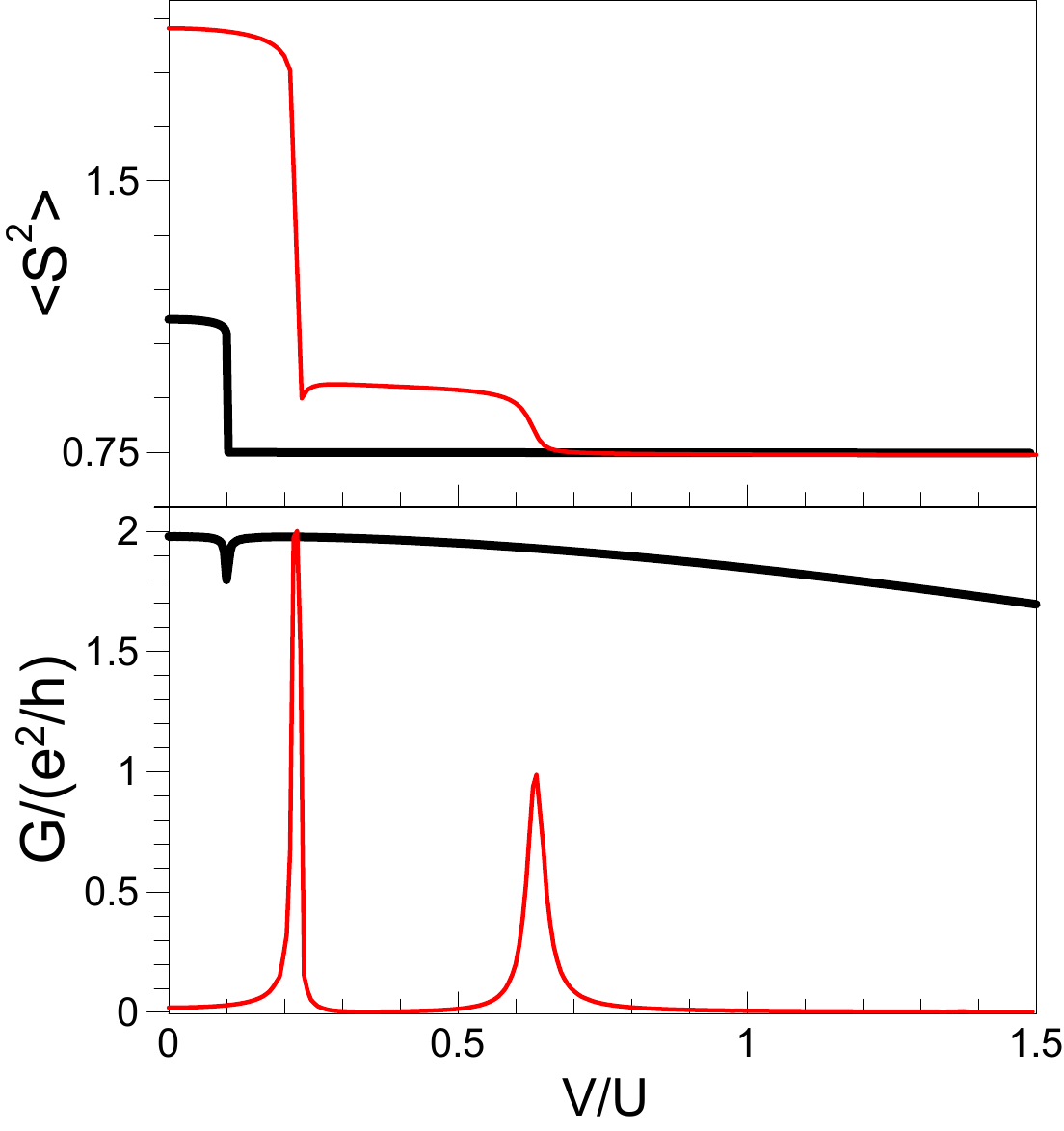}
\caption{(Color online). Bias voltage dependence of the square of the total spin $\langle\mathbf{S}^{2}\rangle$ (upper panel) and the differential conductance $G$ (lower panel) of the double quantum dot system with $t_{LR}=\epsilon=0$, $\gamma=0.9$, and $\Gamma=0.005U$ (thin red line) and $\Gamma=0.5U$ (thick black line).}
\label{S2_G_2}
\end{figure}

In Figure~\ref{S2_G_2} we show the zero-temperature fRG results for the differential conductance $G$ and the average of the square of the total spin $\langle \mathbf{S}^2 \rangle$ as a function of bias voltage $V$, obtained from numerically integrating the non-equilibrium flow equation for the self-energy (Eq.~(\ref{fRG_Eq}) of the main text) for $t_{LR}=\epsilon=0$,  $\gamma=0.9$, and two different choices of $\Gamma/U$. As expected, the DQD system  shows a qualitatively similar picture of the magnetic moment(s) formation in the quantum dots to that of the QQD system. In particular, as for the QQD system, in the double quantum dot system regimes with two, one or none of the magnetic moment(s) in quantum dots can be realized depends on the choice of the geometry of the system $(\Gamma,\gamma)$, for example, for $(\Gamma,\gamma)=(0.005,0.9)$ we have $\langle \mathbf{S_{e/o}}^2 \rangle\approx 3/4$, while for $(\Gamma,\gamma)=(0.5,0.9)$ we find $\langle \mathbf{S_{o}}^2 \rangle\approx 3/4$, $\langle \mathbf{S_{e}}^2 \rangle\approx 3/8$, which corresponds to two and one magnetic moment in the quantum dots at $V=0$, respectively. The possibility of the transition to the state with the local magnetic moment has been studied previously for the DQD system in the equilibrium, in particular, within the Matsubara fRG approach.


As it is seen from Fig.~\ref{S2_G_2}, the application of the bias voltage leads to suppression of the magnetic moment(s) (if they exist at $V=0$) and with increasing of bias voltage the DQD system undergoes evolution from the magnetic moment(s) to the non-magnetic state in such a way, that almost completely corresponds to the one obtained for the QQD system.  Furthermore, the magnetic phase with the fractional value of the magnetic moment also appears when two magnetic moments exist at $V=0$. The differential conductance curves of the DQD system also look similar to those of the QQD system if the corresponding systems have the same magnetic states at $V=0$, however, the conductance of the DQD system shows somewhat different behavior near the phase transition region and does not demonstrate the presence of the negative differential conductance effects for intermediate interactions $U$. 

\section{Landauer-like contribution to conductance}
In this appendix, we show that Eq.~(\ref{G1}) of the main text can be written in a Landauer-like form
\begin{equation}
G_{\sigma}^{0}=\dfrac{2e^{2}}{h}\Gamma_{L}\Gamma_{R}\sum_{\alpha}{\left|\mathcal{G}_{14;\sigma}^{r}\left(\omega=\mu_{\alpha}\right)\right|^{2}}.
\label{G_V_Landauer}
\end{equation}
Using the definitions of the retarded $(p=r)$, advanced $(p=a)$, and Keldysh $(p=K)$ Green functions $\mathcal{G}^{p}$ and the corresponding of self-energies $\Sigma^{p}$, one can write the following identities 
\begin{equation}
\widetilde{\Sigma}^{K}=\widetilde{\Sigma}^{a}-\widetilde{\Sigma}^{r}\\-2\widetilde{\Sigma}^{+-}
\label{SigmaK}
\end{equation}
and
\begin{equation}
\mathcal{G}^{K}=2\mathcal{G}^{+-}-\mathcal{G}^{r}+\mathcal{G}^{a},
\label{GK}
\end{equation}
where we have introduced the notation $\widetilde{\Sigma}^p=\Sigma^p+\Sigma^p_{\rm bath}$; here and below we omit the upper index ;0 of the Green functions and self energies, assuming $\Lambda\rightarrow 0$ limit in the equations of this Appendix.\par
Substituting the Eq.~(\ref{SigmaK}) and Eq.~(\ref{GK}) into the Dyson equation for the Keldysh Green function
\begin{eqnarray}
\mathcal{G}^{K}&=&\left(1+\mathcal{G}^{r}\widetilde{\Sigma}^{r}\right)\mathcal{G}^{K}_{\rm dots}\left(1+\widetilde{\Sigma}^{a}\mathcal{G}^{a}\right)\notag\\&+&\mathcal{G}^{r}\widetilde{\Sigma}^{K}\mathcal{G}^{a}
\end{eqnarray}
one can write the Green function $\mathcal{G}^{+-}$ as
\begin{eqnarray}
\mathcal{G}^{+-}=&-&\dfrac{1}{2}\mathcal{G}^{r}\Bigl[(\mathcal{G}^{r}_{\rm dots})^{-1}-(\mathcal{G}^{a}_{\rm dots})^{-1}\Bigr.\notag\\&-&\Bigl.(\mathcal{G}^{r}_{\rm dots})^{-1}\mathcal{G}^{K}_{\rm dots}(\mathcal{G}^{a}_{\rm dots})^{-1}+2\widetilde{\Sigma}^{+-}\Bigr]\mathcal{G}^{a},
\label{Gpmv1}
\end{eqnarray}
where $\mathcal{G}^{p}_{\rm dots}$ $(p=r,a,K)$ are the Green functions for $\Sigma^{p}=0$ and $\Sigma^{p}_{\rm bath}=0$.\par
Taking into account the explicit form of the Green function (see Eq.~(\ref{Gf}) in the main text), we obtain
\begin{eqnarray}
&&(\mathcal{G}^{r}_{\rm dots})^{-1}-(\mathcal{G}^{a}_{\rm dots})^{-1}\notag\\&&=(\mathcal{G}^{r}_{\rm dots})^{-1}\left(\mathcal{G}^{a}_{\rm dots}-\mathcal{G}^{r}_{\rm dots}\right)(\mathcal{G}^{a}_{\rm dots})^{-1}\notag\\&&=
(\mathcal{G}^{r}_{\rm dots})^{-1}\left(\mathcal{G}^{-+}_{\rm dots}-\mathcal{G}^{+-}_{\rm dots}\right)(\mathcal{G}^{a}_{\rm dots})^{-1}=0
\end{eqnarray}
and 
\begin{equation}
\mathcal{G}^{K}_{\rm dots}=\mathcal{G}^{+-}_{\rm dots}+\mathcal{G}^{-+}_{\rm dots}=0.
\end{equation}
Therefore, Eq.~(\ref{Gpmv1}) reduces to  
\begin{equation}
\mathcal{G}^{+-}=-\mathcal{G}^{r}\Sigma_{\rm bath}^{+-}\mathcal{G}^{a},
\end{equation}
where we have exploited the fact that in our approach $\Sigma^{+-}=0$.\par
Then, using the expression for $\Sigma_{\rm bath}^{+-}$, given by Eq.~(\ref{Sigma_bath}) of the main text, we can write the diagonal elements of $\mathcal{G}^{+-}$ as
\begin{eqnarray}
\mathcal{G}^{+-}_{jj;\sigma}=&-&i\Gamma_{L}\left(1+\sgn(\omega-\mu_{L})\right)\left|\mathcal{G}^{r}_{j1;\sigma}\right|^{2}\notag\\&-&i\Gamma_{R}\left(1+\sgn(\omega-\mu_{R})\right)\left|\mathcal{G}^{r}_{j4;\sigma}\right|^{2},
\end{eqnarray}
where we have used that $\mathcal{G}^{a}=(\mathcal{G}^{r})^{\dagger}$.\par
Finally, substitution of these results into Eq.~(\ref{G1}) of the main text leads to Eq.~(\ref{G_V_Landauer}).



\clearpage
\newpage
\end{document}